\documentclass[11pt,a4paper]{article}

\usepackage{jcappub}
\usepackage{aas_macros}

\newcommand{\be}{\begin{equation}}
\newcommand{\ee}{\end{equation}}           
\newcommand{\fnl}{f_{\rm{NL}}}
\newcommand{\gnl}{g_{\rm{NL}}}
\newcommand{\taunl}{\tau_{\rm{NL}}}

\begin{document}

\setcounter{page}{0}

\title{Local non-Gaussianity from rapidly varying sound speeds}

\author{Jon Emery,}
\author{Gianmassimo Tasinato}
\author{and David Wands}

\affiliation{Institute of Cosmology \& Gravitation, University of Portsmouth,}
\affiliation{Dennis Sciama Building, Portsmouth, PO1 3FX, United Kingdom}

\emailAdd{jon.emery@port.ac.uk}
\emailAdd{gianmassimo.tasinato@port.ac.uk}
\emailAdd{david.wands@port.ac.uk}

\abstract{We study the effect of non-trivial sound speeds on local-type non-Gaussianity during multiple-field inflation. To this end, we consider a multiple-DBI model and use the $\delta N$ formalism to track the super-horizon evolution of perturbations. By adopting a sum separable Hubble parameter we derive analytic expressions for the relevant quantities in the two-field case, valid beyond slow variation. We find that non-trivial sound speeds can, in principle, curve the trajectory in such a way that significant local-type non-Gaussianity is produced. Deviations from slow variation, such as rapidly varying sound speeds, enhance this effect. To illustrate our results we consider two-field inflation in the tip regions of two warped throats and find large local-type non-Gaussianity produced towards the end of the inflationary process.}


\notoc

\maketitle

\newpage

\setcounter{page}{1}

\section{Introduction\label{sec:introduction}}

Whilst inflation provides a compelling mechanism for producing the initial conditions on which the hot big bang relies, a consistent model proves elusive (see \cite{Lyth99,Lyth09,Mazumdar10} for reviews).  This is due, in part, to the limited information available from the two-point statistics of primordial density perturbations. Interest has therefore focused on potential non-Gaussian signatures to discern between otherwise degenerate models \cite{Komatsu09}, particularly since the sensitivity of forthcoming observations is set to improve by at least an order of magnitude \cite{Planck06}. As such, it is important to understand the correspondence between inflationary dynamics and the resultant form of non-Gaussianity (see \cite{Bartolo04,Chen10} for recent reviews).

Inflation is capable of producing non-Gaussianity in a variety of ways. For example, the conversion between entropy and adiabatic modes \emph{during}\footnote{Alternatively, non-linearities can develop through the conversion between entropy and adiabatic modes \emph{after} inflation. For example, the curvaton mechanism \cite{Moroi01,Lyth02,Lyth03} and modulated reheating \cite{Kofman03,Dvali04}. See
\cite{Burgess10,Cicoli12} for recent realisations of such scenarios in string theory.} multiple-field inflation \cite{Gordon01,Nibbelink02,Rigopoulos04} allows the curvature perturbation $\zeta$ to evolve on super-horizon scales \cite{Wands00,Lyth05}. Such non-linearities  can, in certain circumstances, produce local-type non-Gaussianity during inflation \cite{Bernardeau02,Vernizzi06,Rigopoulos06,Rigopoulos07,Yokoyama07,Battefeld07,Yokoyama08,Byrnes08,Sasaki08,Byrnes09,Battefeld09,Wang10,Tzavara11,Elliston11a,Elliston11b,Meyers11-a,Meyers11-b,Peterson11a,Watanabe11,Choi12,Mazumdar12,Frazer11a,Battefeld12}. Some simplifying assumptions are usually made to make such calculations analytically tractable however, owing to the increased phase space. For example, Vernizzi \& Wands \cite{Vernizzi06} calculated the non-linearity parameter in canonical two-field models possessing a sum separable potential $V(\phi,\chi)=V^{(\phi)}(\phi)+V^{(\chi)}(\chi)$. Byrnes \& Tasinato \cite{Byrnes09} extended this beyond slow roll by exploiting the Hamilton-Jacobi formalism \cite{Salopek90,Kinney97} and a sum separable Hubble parameter $H(\phi,\chi)=H^{(\phi)}(\phi)+H^{(\chi)}(\chi)$. The broad conclusion is that a turn in the trajectory, caused by the potential, can produce significant local-type non-Gaussianity during inflation. Furthermore, the effect is enhanced by violations of slow roll.

Single field models with non-standard kinetic terms, often motivated by string theory, provide an alternative source of non-Gaussianity (see \cite{Chen07,Koyama10} and references therein). The models we are concerned with have a characteristic sound speed $c_{s}$, where $c_{s}=1$ in the canonical case.\footnote{Whilst commonly referred to as the sound speed, it should be noted that this is technically the phase speed of fluctuations. See \cite{Christopherson09} for further discussion on this point.} Non-Gaussianity is caused by quantum field interactions on sub-horizon scales and is typically of equilateral form. In the Dirac-Born-Infeld (DBI) case \cite{Silverstein04,Alishahiha04}, for example, a probe D-brane moving along the radial direction of a warped throat causes inflation. Here the radial co-ordinate of the D-brane plays the role of the inflaton. In this case the sound speed acts as an effective speed limit, facilitating inflation on a steep potential and producing significant equilateral type non-Gaussianity. 

In more general models it is expected that both contributions to non-Gaussianity will be relevant. The combined effect of non-standard kinetic terms and multiple-field dynamics has been considered by \cite{Langlois08,Langlois08b,Arroja08,RenauxPetel09a,Gao09,Mizuno09,Langlois09a,Gao09a,Cai09,Cai09-1,Pi11}, with emphasis generally on the effect of multiple-field dynamics on equilateral type non-Gaussianity. Here however, we focus on the effect of non-canonical kinetic terms on local-type non-Gaussianity, produced by multiple dynamical fields (see \cite{RenauxPetel09,Dias12,Khosravi12} for alternative examples). Specifically, we consider whether non-trivial sound speeds, as opposed to the potential, can produce local-type non-Gaussianity during a turn in the trajectory. 

It should be noted that, in the canonical case, the effect is enhanced by deviations from slow roll. It is reasonable to expect similar behaviour in the scenario considered here, corresponding to rapidly varying sound speeds. Such behaviour has been explored by \cite{Khoury09,Noller11,Nakashima11,Ribeiro12,Park12} in the single field case, who find additional features in the equilateral limit of the bispectrum. In this paper however, we show that rapidly varying sound speeds in multiple-field models can lead to large non-Gaussianity of local form.

In this work we choose a multiple-DBI model, akin to that of \cite{Cai09,Cai09-1,Pi11}, as a concrete example of multi-component inflation with non-standard kinetic terms. We then employ the $\delta N$ formalism to track the super-horizon evolution of perturbations using the field fluctuations at horizon exit and the subsequent background trajectory. The trajectory is a solution of the homogeneous equations of motion, which we write in Hamilton-Jacobi form and derive in an appendix. Whilst the dynamics are described more naturally in this form, it also allows the adoption of a sum separable Hubble parameter, as in \cite{Byrnes09}, to treat the two-field case both analytically and beyond slow variation. With these expressions, we ascertain whether non-trivial sound speeds can produce a turn in the trajectory during which local-type non-Gaussianity is produced. To illustrate our results we consider inflation in the tip regions of two warped throats, where one expects the sound speeds to increase rapidly and violate slow variation.

The outline of the paper is as follows. We begin in section~\ref{sec:model} by introducing the multiple-DBI model, before introducing and reformulating the relevant quantities using the $\delta N$ formalism. By utilising a sum separable Hubble parameter we analyse these expressions analytically in section~\ref{sec:separable}, before summarising the necessary conditions for large local-type non-Gaussianity. We illustrate these results in section~\ref{sec:example} by studying inflation in the tip regions of two warped throats. Finally, we conclude in section~\ref{sec:conclusions}.

Throughout this paper we use the $(-,+,+,+)$ metric signature and set $M_{\textrm{P}}=c=1$, where $M_{\textrm{P}}=\frac{1}{\sqrt{8\pi G}}$ is the reduced Planck mass. Capital latin indices label scalar fields and any summation is explicit. Greek indices label space-time co-ordinates whilst lower case latin indices label spatial co-ordinates only, where the Einstein summation convention is adopted. Finally, commas denote partial derivatives and over-dots represent derivatives with respect to time. 

\section{Multiple-DBI inflation, non-Gaussianity and the $\delta N$ formalism\label{sec:model}}

We begin this section by introducing a multiple-DBI model as a specific realisation of multi-component inflation with non-standard kinetic terms. With this model in mind we introduce the relevant quantities, paying particular attention to non-Gaussianity. Finally, we recast these expressions with the $\delta N$ formalism to study their evolution on super-horizon scales, of which we are primarily interested.   

\newpage

\subsection{Multiple-DBI inflation\label{sec:model-homogeneous}}

We begin by considering the following action

      \be S=\frac{1}{2} \int d^{4}x\sqrt{-g}\bigl[R+2\sum_{I}P_{I}-2V\bigr],\label{eq:action-general}\ee

\noindent where $P_{I}$ is a function of the single scalar field $\phi_{I}$ and kinetic function $X_{I}=-\frac{1}{2}g^{\mu\nu}\phi_{I,\mu}\phi_{I,\nu}$, whereas the potential $V$ is a function of the set of scalar fields $\phi=\{ \phi_{1},\phi_{2},...,\phi_{N}\}$. For the spatially flat Friedmann-Robertson-Walker (FRW) metric the collection of scalar fields act as a perfect fluid with isotropic pressure $P=\sum_{I}P_{I}-V$ and energy density $\rho=\sum_{I}\rho_{I}+V$, where $\rho_{I}=2X_{I}P_{I,X_{I}}-P_{I}$. In appendix~\ref{sec:appendix} we derive the corresponding homogeneous equations of motion by recasting the action (\ref{eq:action-general}) using the Arnowitt-Deser-Misner (ADM) formalism \cite{Arnowitt62}. This ensures that the results are in Hamilton-Jacobi form \cite{Salopek90,Kinney97} in which the Hubble parameter is a function of the scalar fields and takes precedence over the potential. Not only is this more suited to the case of non-trivial sound speeds but it will also prove useful for studying violations of slow variation. The resultant equations of motion are
	    
      \vspace{-15pt}
      \begin{align}
      \dot{\phi}_{I}&=-\frac{2}{P_{I,X_{I}}}H_{,I},\label{eq:field-general}\\[15pt] 
      3H^{2}&=\sum_{I}\left(\frac{4H_{,I}^{2}}{P_{I,X_{I}}}-P_{I}\right)+V,\label{eq:friedman-general} 
      \end{align}

\noindent where $H$ is the Hubble parameter and is, in general, a function of the set of scalar fields~$\phi$. Furthermore, it is convenient to define individual sound speeds 
      
      \vspace{2pt}
      \be c^{(I)}=\sqrt{\frac{P_{I,X_{I}}}{\rho_{I,X_{I}}}}=\sqrt{\frac{P_{I,X_{I}}}{P_{I,X_{I}}+2X_{I}P_{I,X_{I}X_{I}}}}.\label{eq:sound-speed-general}\ee
      \vspace{2pt}

\noindent We now model inflation as driven by $N$ probe D3 branes traversing $N$ distinct warped throats glued to a compact Calabi-Yau manifold in type IIB string theory.\footnote{This model is then distinct from the first example of multiple-field DBI \cite{Easson08}, in which a single brane descends a warped throat along radial and angular coordinates.} This was first considered by \cite{Cai09,Cai09-1,Pi11} to study the effect of multiple sound horizons on equilateral type non-Gaussianity. The corresponding expression for $P_{I}$ is

      \be P_{I}=\frac{1}{f^{(I)}}\left(1-\sqrt{1-2f^{(I)}X_{I}}\right).\ee

\noindent Here $f^{(I)}$ parameterises the warped brane tension of throat $I$ and is a function of $\phi_{I}$ only. \newpage \noindent On substitution into (\ref{eq:field-general})--(\ref{eq:sound-speed-general}) we find the following equations of motion 

      \vspace{-10pt}
      \begin{align}	
      \dot{\phi}_{I}&=-2c^{(I)}H_{,I},\label{eq:field-dbi}\\[15pt]
      3H^{2}&=V-\sum_{I}\frac{1}{f^{(I)}}\left(1-\frac{1}{c^{(I)}}\right),\label{eq:friedman-dbi} 
      \end{align}
      \vspace{-5pt}

\noindent where $c^{(I)}$ is given by
	    
      \vspace{-2pt}
      \be c^{(I)}=\frac{1}{\sqrt{1+4f^{(I)}H_{,I}^{2}}},\label{eq:sound-speed-dbi}\ee
      \vspace{2pt}

\noindent and we have used (\ref{eq:field-general}) to eliminate $X_{I}$ in the above. As a result, $f^{(I)}$ remains a function of $\phi_{I}$ whilst $H$, $c^{(I)}$, $V$ and $\dot{\phi_{I}}$ generally depend on the collection of fields $\phi$. Finally, we define the following slow variation parameters:
	    
      \vspace{-10pt}
      \begin{align}
      \epsilon&=-\frac{\dot{H}}{H^{2}}=\sum_{I}\epsilon^{(I)}=\sum_{I}2c^{(I)}\left(\frac{H_{,I}}{H}\right)^{2},\nonumber\\[10pt]
      \eta^{(I)}&=\sum_{J}\eta^{(IJ)}=\sum_{J}2c^{(J)}\frac{H_{,J}}{H_{,I}}\frac{H_{,IJ}}{H},\nonumber\\[10pt]
      s^{(I)}&=-\frac{\dot{c}^{(I)}}{Hc^{(I)}}=\sum_{J}s^{(IJ)}=\sum_{J}2c_{,J}^{(I)}\frac{c^{(J)}}{c^{(I)}}\frac{H_{,J}}{H}.	
      \end{align}

\noindent The only general requirement on these parameters is $\epsilon<1$, corresponding to $\ddot{a}>0$, where $a$ is the scale factor. We shall state explicitly when the additional restriction of slow variation is required, which corresponds to $\epsilon^{(I)},\eta^{(I)},s^{(I)}\ll1$.   

\subsection{Perturbations and non-Gaussianity\label{sec:model-perturbations}}

It is convenient to introduce the primordial curvature perturbation on uniform density hypersurfaces $\zeta(t,x^{i})$, to characterise the scalar degree of freedom in the primordial perturbations (see \cite{Malik01,Malik09} for explicit definitions). The power spectrum $P_{\zeta}$ is then defined using the two-point function 
	    
      \vspace{-5pt}
      \be \langle\zeta_{\mathbf{k_{1}}}\zeta_{\mathbf{k_{2}}}\rangle = (2\pi)^{3} P_{\zeta}(k_{1}) \, \delta^{3}(\mathbf{k_{1}}+\mathbf{k_{2}}),\label{eq:two-point-function}\ee
      \vspace{2pt}
	    
\noindent where $\zeta_{\mathbf{k}}$ is the Fourier transform of $\zeta$, $\mathbf{k_{i}}$ are comoving wavevectors and $\delta^{3}$ is the three dimensional Dirac delta function. Moreover, the dimensionless power spectrum $\mathcal{P}_{\zeta}(k)$ is often defined, related to the power spectrum by $\mathcal{P}_{\zeta}=\frac{k^{3}}{2\pi^{2}}P_{\zeta}$. Given this, the spectral index $n_{\zeta}$  characterises the deviation of $\mathcal{P}_{\zeta}$ from scale invariance

      \vspace{-3pt}
      \be n_{\zeta}-1=\frac{d\log\mathcal{P}_{\zeta}{}}{d\log k},\ee
      \vspace{-3pt}

 \noindent where scale invariance corresponds to $n_{\zeta}=1$. The two-point function completely defines the statistics of the field if $\zeta$ is purely Gaussian. Any signature of non-Gaussianity will be encoded in the connected contributions to higher order correlators, the next order being the three-point function

      \vspace{-3pt}
      \be \langle \zeta_{\mathbf{k_{1}}}\zeta_{\mathbf{k_{2}}}\zeta_{\mathbf{k_{3}}} \rangle = (2\pi)^{3}B_{\zeta}(k_{1},k_{2},k_{3})\delta^{3}(\mathbf{k_{1}}+\mathbf{k_{2}}+\mathbf{k_{3}}),\label{eq:three-point-function}\ee
      \vspace{-3pt}

\noindent where the bispectrum $B_{\zeta}$ is analogous to the power spectrum. We parameterise the deviation from Gaussianity by taking the ratio of the bispectrum to a combination of the power spectra
	    
      \vspace{4pt}
      \be \frac{6}{5}\fnl(k_{1},k_{2},k_{3})=\frac{B_{\zeta}(k_{1},k_{2},k_{3})}{P_{\zeta}(k_{1})P_{\zeta}(k_{2}) + P_{\zeta}(k_{1})P_{\zeta}(k_{3}) + P_{\zeta}(k_{2})P_{\zeta}(k_{3})},\ee
      \vspace{4pt}

\noindent where $\fnl$ is the $k$-dependent non-linearity parameter\footnote{There are a number of sign conventions in the literature and we adopt those of \cite{Byrnes09} for ease of comparison. See \cite{Wands10} for a summary of the various conventions.} and the numerical coefficient is a relic of the original definition in terms of the Bardeen potential $\Phi=\frac{3}{5}\zeta$. Assuming a scale invariant dimensionless power spectrum and using the relation $\mathcal{P}_{\zeta}=\frac{k^{3}}{2\pi^{2}}P_{\zeta}(k)$, the non-linearity parameter can be re-written as
	      
      \vspace{4pt}
      \be  \frac{6}{5}\fnl(k_{1},k_{2},k_{3})=\frac{\prod_{i}k_{i}^{3}}{\sum_{i}k_{i}^{3}}\frac{B_{\zeta}(k_{1},k_{2},k_{3})}{4\pi^{4}\mathcal{P}_{\zeta}^{2}},\label{eq:fnl}\ee
      \vspace{4pt}

\noindent which is a more useful form for the following section. 

\subsection{Super-horizon evolution and the $\delta N$ formalism\label{sec:model-deltaN}}

The conversion of entropy to adiabatic modes during multiple-field inflation \cite{Gordon01,Nibbelink02,Rigopoulos04} allows the curvature perturbation $\zeta$ to evolve on super-horizon scales \cite{Wands00,Lyth05}, and it is on this evolution we wish to focus. As such, we adopt the $\delta N$ formalism \cite{Starobinskivi85,Sasaki96,Sasaki98,Wands00,Lyth05-1}, a powerful technique for evolving $\zeta$ on super-horizon scales using only the field fluctuations at horizon exit and the homogeneous field evolution thereafter. 

\newpage

In order to employ the $\delta N$ formalism we make two restrictions on the background dynamics. Since our scenario admits multiple sound horizons we demand that these are comparable whilst observable scales exit during inflation, such that $c^{(I)} = c_{\star}$ for all $I$ during this interval.\footnote{See \cite{Cai09-1,Pi11} for a discussion of when this is not the case.} Horizon exit\footnote{For brevity we will refer to the `sound-horizon' simply as the `horizon'. There should not be any ambiguity since, in this context, the sound-horizon will be the only relevant scale.} therefore equates to evaluating a quantity when $c_{\star}k = a_{\star}H_{\star}$. Our second restriction is that of slow variation at horizon exit, since this simplifies the spectrum of field fluctuations produced at horizon exit.

Given these restrictions, the $\delta N$ formalism uses the separate Universe approach \cite{Wands00,Sasaki96,Sasaki98,Rigopoulos03} to allow the identification of the curvature perturbation $\zeta$ with the difference in the number of e-folds between the perturbed ($N$) and homogeneous background ($N_{0}$) universes, evaluated between an initially flat hypersurface $t_{\star}$ (e.g. shortly after horizon exit) and a final uniform density hypersurface $t_{f}$ (e.g. early in the radiation dominated epoch)

      \be \zeta(t_{f},x^{i})=\delta N(t_{f},x^{i}) = N(t_{\star},t_{f},x^{i}) - N_{0}(t_{\star},t_{f}).\ee
      \vspace{1pt}

\noindent  Assuming $N$ is a local function of the scalar fields at horizon exit, which are split into a homogeneous background and local perturbation $\phi^{I}(t_{\star},x^{i})=\phi^{I}(t_{\star})+\delta\phi^{I}(t_{\star},x^{i})$, $\zeta$ can be expanded in powers of $\delta\phi^{I}$ as
	    
      \vspace{-7pt}
      \begin{align}
      \zeta &=  \sum_{I}N_{,I}\delta\phi^{I} + \frac{1}{2}\sum_{IJ}N_{,IJ}\left( \delta\phi^{I}\delta\phi^{J} - \langle \delta\phi^{I}\delta\phi^{J} \rangle  \right)+ \ldots,\label{eq:deltaN-expansion}
      \end{align}
      \vspace{2pt}

\noindent where $N_{,I}$ is with respect to the field $I$ at horizon exit. Using this expansion, we are able calculate the relevant quantities (e.g. $\fnl$) at time $t_{f}$ given the field fluctuations at time $t_{\star}$ and the homogeneous field evolution between $t_{\star}$ and $t_{f}$. For example, on substitution of the expansion (\ref{eq:deltaN-expansion}) into the two-point function (\ref{eq:two-point-function}), one finds that the dimensionless power spectrum can be expressed as \cite{Lyth05-1,Byrnes06}

      \be \mathcal{P}_{\zeta}=\sum_{I}N^{2}_{,I}\mathcal{P}_{\star}.\label{eq:power-spectrum-deltaN}\ee	
      \vspace{2pt}

\noindent Here we have defined the dimensionless power spectrum of scalar field fluctuations at horizon exit using the two point function

      \be \langle \delta\phi^{I}_{\mathbf{k_{1}}} \delta \phi^{J}_{\mathbf{k_{2}}}\rangle=(2\pi)^{3}\,\delta^{IJ}\frac{2\pi^{2}}{k_{1}^{3}}\mathcal{P}_{\star}\,\delta^{3}(\mathbf{k_{1}}+\mathbf{k_{2}}),\hspace{20pt}\mathcal{P}_{\star}=\left(\frac{H_{\star}}{2\pi}\right)^{2},\ee
      
\newpage

\noindent where we have used slow variation at horizon exit and $\delta^{IJ}$ is the kronecker delta symbol. The spectral index can be treated similarly, giving  

      \be n_{\zeta}-1=-2\epsilon_{\star}+\frac{2}{H}\frac{\sum_{IJ}\dot{\phi}_{J}N_{,I}N_{,IJ}}{\sum_{K}N_{,K}^{2}}.\label{eq:spectral-index-deltaN}\ee
      \vspace{3pt}

\noindent Progressing to the three-point function, substitution of the expansion (\ref{eq:deltaN-expansion}) into (\ref{eq:three-point-function}) neatly highlights the physical sources of non-Gaussianity 
	    
      \vspace{-6pt}
      \begin{align}
      \langle \zeta_{\mathbf{k_{1}}}\zeta_{\mathbf{k_{2}}}\zeta_{\mathbf{k_{3}}} \rangle =  \sum_{IJK}&N_{,I}N_{,J}N_{,K}{}\langle\delta\phi^{I}_{\mathbf{k_{1}}}\delta\phi^{J}_{\mathbf{k_{2}}}\delta\phi^{K}_{\mathbf{k_{3}}}\rangle \, +  \nonumber \\[5pt]  
      &\biggl(\,\frac{1}{2}\sum_{IJKL}N_{,I}N_{,J}N_{,KL}\langle\delta\phi^{I}_{\mathbf{k_{1}}}\delta\phi^{J}_{\mathbf{k_{2}}}(\delta\phi^{K}\star\delta\phi^{L})_{\mathbf{k_{3}}}\rangle + 2\,\mathrm{perms}\,\biggr)\label{eq:three-point-function-deltaN},
      \end{align}
      \vspace{2pt}

\noindent where here $\star$ denotes a convolution and `perms' denotes cyclic permutations over the momenta. The expression (\ref{eq:three-point-function-deltaN}) shows two distinct contributions to the three-point function. The first term contributes to non-Gaussianity in $\zeta$ through the intrinsic non-Gaussianity of $\delta\phi^{I}$, produced by quantum field interactions on sub-horizon scales. Whilst this term vanishes identically for Gaussian $\delta\phi^{I}$, $\zeta$ can still be non-Gaussian through the second term. This contribution is due to non-linear behaviour in $\zeta$ on super-horizon scales and is often referred to as local type.\footnote{See \cite{Babich04} for a discussion of the shape dependence of the bispectrum.} It is on this contribution that we shall predominantly focus. 

\noindent Neglecting the connected part of the four-point function and using Wick's theorem to rewrite the four-point functions as products of two-point functions, the latter term can be expressed as \cite{Lyth05-1,Byrnes06} 
	    
      \vspace{-15pt}
      \begin{align}
      \frac{1}{2}\sum_{IJKL}N_{,I}&N_{,J}N_{,KL}\langle\delta\phi^{I}_{\mathbf{k_{1}}}\delta\phi^{J}_{\mathbf{k_{2}}}(\delta\phi^{K}\star\delta\phi^{L})_{\mathbf{k_{3}}}\rangle + 2\,\mathrm{perms}=\nonumber \\[10pt]&(2\pi)^{3}\,4\pi^{4}\mathcal{P}_{\star}^{2}\,\frac{\sum_{i}k_{i}^{3}}{\prod_{i}k_{i}^{3}}\sum_{IJ}N_{,I}N_{,J}N_{,IJ}\,\delta^{3}({\mathbf{k_{1}}+\mathbf{k_{2}}+\mathbf{k_{3}}}),
      \end{align}
      \vspace{2pt}

\noindent such that the bispectrum becomes
	    
      \vspace{-10pt}
      \begin{align}
      B_{\zeta}(k_{1},k_{2},k_{3})=4\pi^{4}\mathcal{P}_{\zeta}^{2}\frac{\sum_{i}k_{i}^{3}}{\prod_{i}k_{i}^{3}}\left(\frac{6}{5}\fnl^{(3)}(k_{1},k_{2},k_{3})+\frac{\sum_{IJ}N_{,I}N_{,J}N_{,IJ}}{(\sum_{K}N_{,K}^{2})^{2}}\right),
      \end{align}

\newpage

\noindent and we have used (\ref{eq:power-spectrum-deltaN}) to replace $\mathcal{P}_{\star}$ with $\mathcal{P}_{\zeta}$. Adopting the notation of \cite{Vernizzi06}, the $k$-dependent parameter $\fnl^{(3)}(k_{1},k_{2},k_{3})$ accounts for the sub-horizon contribution for which we omit an explicit expression, since we are primarily concerned with the super-horizon contribution. We will return to $\fnl^{(3)}$ in our conclusions. Finally, on substitution into (\ref{eq:fnl}), we arrive at

      \vspace{2pt}
      \be \fnl(k_{1},k_{2},k_{3})=\fnl^{(3)}(k_{1},k_{2},k_{3}) + \fnl^{(4)}.\ee
      \vspace{2pt}

\noindent Again adopting the notation of \cite{Vernizzi06}, the $k$-independent parameter\footnote{See \cite{Byrnes10a,Byrnes10b} for scale dependent $\fnl$.} $\fnl^{(4)}$ accounts for the super-horizon contribution and is given by

      \be \frac{6}{5}\fnl^{(4)}=\frac{\sum_{IJ}N_{,I}N_{,J}N_{,IJ}}{(\sum_{K}N_{,K}^{2})^{2}}.\label{eq:fnl4}\ee
      \vspace{5pt}

\noindent Our aim then is to calculate $\mathcal{P}_{\zeta}$, $n_{\zeta}$ and $\fnl^{(4)}$ in the model described above using the expressions (\ref{eq:power-spectrum-deltaN}), (\ref{eq:spectral-index-deltaN}) and (\ref{eq:fnl4}) respectively. To evaluate these expressions we must evaluate the terms $N_{,I}$ and $N_{,IJ}$. Unlike single field inflation however, there is no unique attractor in the multiple-field case \cite{GarciaBellido96}. There are therefore an infinite number of trajectories and, as such, the derivatives of $N$ become non-trivial. One is usually limited to numerical studies without a further restriction and so, to make analytical progress, we focus on scenarios satisfying a sum separable Hubble parameter in the following section.  

\section{The role of rapidly varying sound speeds\label{sec:separable}}

To investigate how local-type non-Gaussianity is affected by rapidly varying sound speeds, we apply the preceding results to a subset of analytically tractable cases. We note again that such behaviour was explored in the single field scenario by \cite{Khoury09,Noller11,Nakashima11,Ribeiro12,Park12}, in the context of equilateral type non-Gaussianity. In this section however, we show that rapidly varying sound speeds in the multiple-field case can, in principle, lead to large local-type non-Gaussianity.  

Given the lack of a unique attractor in multiple-field scenarios, certain restrictions are required to evaluate the expressions in section~\ref{sec:model-deltaN} analytically.  For example, \cite{Vernizzi06} calculated the non-linearity parameter analytically for canonical two-field models by demanding a sum separable potential $V(\phi,\chi)=V^{(\phi)}(\phi)+V^{(\chi)}(\chi)$. This invoked the slow roll approximation throughout however, to reduce the second order field equations to first order expressions. To study more diverse dynamics, \cite{Byrnes09} developed an analogous formalism by writing the field equations in Hamilton-Jacobi form \cite{Salopek90,Kinney97}, in which they are automatically first order, and considered a sum separable Hubble parameter $H(\phi,\chi)=H^{(\phi)}(\phi)+H^{(\chi)}(\chi)$. As a result, the slow roll approximation can be discarded after horizon exit and one can reliably study regimes in which slow roll is violated. 

\newpage

In this section we adopt the method invoked by \cite{Byrnes09} and demand a sum separable Hubble parameter. Whilst allowing the violation of slow variation after horizon exit, this also suits the case of non-standard kinetic terms, since the dynamics are more naturally described in the Hamilton-Jacobi formalism (see section~\ref{sec:model-homogeneous}). Given this restriction, we exploit the resultant integral of motion to derive analytic expressions for the derivatives of $N$ and in turn $\mathcal{P}_{\zeta}$, $n_{\zeta}$ and $\fnl^{(4)}$. The calculation closely follows \cite{Byrnes09} and we adopt analogous notation to clearly illustrate the additional effects of non-standard kinetic terms.  

\subsection{Two-field model with a sum separable Hubble parameter\label{sec:separable-two-field}}

We begin by considering the two-field case, with fields $\phi$ and $\chi$, and demand a sum separable Hubble parameter
     
      \be H(\phi,\chi)=H^{(\phi)}(\phi)+H^{(\chi)}(\chi),\label{eq:separable-Hubble}\ee
      \vspace{1pt}

\noindent which leads to a number of simplifications. From equation (\ref{eq:sound-speed-dbi}), the sound speed $c^{(I)}$ becomes a function of its respective field $\phi_{I}$ only, such that $c^{(\phi)}(\phi)$ and $c^{(\chi)}(\chi)$. Moreover, cross derivatives of $H$ (i.e. $H_{,\phi\chi}$) become zero. The combined effect is to reduce the number of relevant slow variation parameters
	
      \vspace{-3pt}
      \begin{align}
      \epsilon^{(\phi)}&=2c^{(\phi)}\left(\frac{H_{,\phi}}{H}\right)^{2},\hspace{35pt}
      \epsilon^{(\chi)}=2c^{(\chi)}\left(\frac{H_{,\chi}}{H}\right)^{2},\nonumber\\[13pt]
      \eta^{(\phi)}&=2c^{(\phi)}\frac{H_{,\phi\phi}}{H},\hspace{53pt}
      \eta^{(\chi)}=2c^{(\chi)}\frac{H_{,\chi\chi}}{H},\nonumber\\[13pt]
      s^{(\phi)}&=2c^{(\phi)}_{,\phi}\frac{H_{,\phi}}{H},\hspace{58pt}
      s^{(\chi)}=2c^{(\chi)}_{,\chi}\frac{H_{,\chi}}{H},\label{eq:slow-variation-separable}
      \end{align}
      \vspace{0pt}

\noindent where $\epsilon=\epsilon^{(\phi)}+\epsilon^{(\chi)}$ and we again emphasise that $\eta^{(I)}$ and $s^{(I)}$ are not constrained to be small after horizon exit. The most crucial simplification however, is the ability to calculate the derivatives in section~\ref{sec:model-deltaN} analytically. To this end, we use (\ref{eq:separable-Hubble}) to write the number of e-folds between $t_{\star}$ and $t_{f}$ as

      \be N(t_{\star},t_{f})=\int_{t_{\star}}^{t_{f}}Hdt=
      \frac{1}{2}\int_{\phi_{f}}^{\phi_{\star}}\frac{H^{(\phi)}}{c^{(\phi)}H^{(\phi)}_{,\phi}}d\phi+
      \frac{1}{2}\int_{\chi_{f}}^{\chi_{\star}}\frac{H^{(\chi)}}{c^{(\chi)}H^{(\chi)}_{,\chi}}d\chi,\label{eq:e-folds}\ee
      \vspace{2pt}
											
\noindent which describes the evolution along the classical trajectory. Moreover, (\ref{eq:separable-Hubble}) permits the following dimensionless integral of motion

      \vspace{5pt}
      \be \mathcal{C}=-\int\frac{d\phi}{c^{(\phi)}H^{(\phi)}_{,\phi}}+\int\frac{d\chi}{c^{(\chi)}H^{(\chi)}_{,\chi}},\label{eq:C}\ee

\noindent  which is constant and unique for each classical trajectory, characterising motion orthogonal to it. By combining (\ref{eq:e-folds}) and (\ref{eq:C}), we find the following expressions for the first derivatives of $N$ 
	  
      \vspace{-15pt}
      \begin{align}	
      N_{,\phi_{\star}}&=\frac{1}{2}\left(\frac{H^{(\phi)}_{\star}}{c^{(\phi)}_{\star}H^{(\phi)}_{,\phi_{\star}}}
      -\frac{H^{(\phi)}_{f}}{c^{(\phi)}_{f}H^{(\phi)}_{,\phi_{f}}}\frac{\partial\phi_{f}}{\partial\phi_{\star}}
      -\frac{H^{(\chi)}_{f}}{c^{(\chi)}_{f}H^{(\chi)}_{,\chi_{f}}}\frac{\partial\chi_{f}}{\partial\phi_{\star}}\right),
      \nonumber\\[20pt]
      N_{,\chi_{\star}}&=\frac{1}{2}\left(\frac{H^{(\chi)}_{\star}}{c^{(\chi)}_{\star}H^{(\chi)}_{,\chi_{\star}}}
      -\frac{H^{(\chi)}_{f}}{c^{(\chi)}_{f}H^{(\chi)}_{,\chi_{f}}}\frac{\partial\chi_{f}}{\partial\chi_{\star}}
      -\frac{H^{(\phi)}_{f}}{c^{(\phi)}_{f}H^{(\phi)}_{,\phi_{f}}}\frac{\partial\phi_{f}}{\partial\chi_{\star}}\right),
     \label{eq:dN-dphi}
      \end{align}
      \vspace{5pt}
								
\noindent where, for clarity, we have included explicit references to $t_{f}$ and $t_{\star}$ via the corresponding subscript. The task now is to calculate the derivatives of the fields at $t_{f}$ with respect to the fields at $t_{\star}$, where we can again invoke the integral of motion (\ref{eq:C}).

\subsection{Evaluating the field derivatives\label{sec:separable-derivatives}}

For brevity we focus only on $\frac{\partial\phi_{f}}{\partial\phi_{\star}}$, with the remaining combinations following analogously.  We begin by using the integral of motion (\ref{eq:C}) to write
	      
      \vspace{-2pt}
      \be \frac{\partial\phi_{f}}{\partial\phi_{\star}}=\frac{d\phi_{f}}{d\mathcal{C}}\frac{\partial\mathcal{C}}{\partial\phi_{\star}}.\label{eq:field-calc-1}\ee \vspace{1pt}
	      
\noindent With respect to the first term, we note that since $t_{f}$ is defined as a constant density hypersurface, $H(\phi_{f},\chi_{f})$ is a constant in a spatially flat Universe.  As a result 

      \be \frac{d\phi_{f}}{d\mathcal{C}}H^{(\phi)}_{,\phi_{f}}+\frac{d\chi_{f}}{d\mathcal{C}}H^{(\chi)}_{,\chi_{f}}=0.\label{eq:field-calc-2}\ee
      \vspace{2pt}

\noindent Furthermore, differentiating (\ref{eq:C}) along $\mathcal{C}$ gives

      \be -\frac{1}{c^{(\phi)}_{f}H^{(\phi)}_{,\phi_{f}}}\frac{d\phi_{f}}{d\mathcal{C}}+\frac{1}{c^{(\chi)}_{f}H^{(\chi)}_{,\chi_{f}}}\frac{d\chi_{f}}{d\mathcal{C}}=1,\label{field-calc-3}\ee
      \vspace{2pt}

\noindent which, combined with (\ref{eq:field-calc-2}) yields the first term in (\ref{eq:field-calc-1})

      \be \frac{d\phi_{f}}{d\mathcal{C}}=-\frac{1}{H^{(\phi)}_{,\phi_{f}}\left(\frac{1}{c^{(\phi)}_{f}H^{(\phi)^{2}}_{,\phi_{f}}}+\frac{1}{c^{(\chi)}_{f}H^{(\chi)^{2}}_{,\chi_{f}}}\right)}.\label{field-calc-4}\ee
      \vspace{3pt}

\noindent The second term is simply given by the definition of $\mathcal{C}$ 

      \vspace{3pt}
      \be \frac{\partial\mathcal{C}}{\partial\phi_{\star}}=-\frac{1}{c^{(\phi)}_{\star}H^{(\phi)}_{,\phi_{\star}}},\label{eq:field-calc-5}\ee
      \vspace{3pt}

\noindent and we combine (\ref{field-calc-4}) and (\ref{eq:field-calc-5}) to find the result for $\frac{\partial\phi_{f}}{\partial\phi_{\star}}$
    
      \vspace{3pt}
      \be \frac{\partial\phi_{f}}{\partial\phi_{\star}}=\frac{1}{c^{(\phi)}_{\star}H^{(\phi)}_{,\phi_{\star}}H^{(\phi)}_{,\phi_{f}}\left(\frac{1}{c^{(\phi)}_{f}H^{(\phi)^{2}}_{,\phi_{f}}}+\frac{1}{c^{(\chi)}_{f}H^{(\chi)^{2}}_{,\chi_{f}}}\right)}.\ee
      \vspace{3pt}

\noindent Finally, we use the slow variation parameters (\ref{eq:slow-variation-separable}) and include the remaining combinations of derivatives to find 
	      
      \vspace{-14pt}
      \begin{align}
      \frac{\partial\phi_{f}}{\partial\phi_{\star}}&=\frac{H_{f}}{H_{\star}}\,\frac{\epsilon^{(\chi)}_{f}}{\epsilon_{f}}\,
      \left(\frac{c^{(\phi)}_{f}}{c^{(\phi)}_{\star}}\,
      \frac{\epsilon^{(\phi)}_{f}}{\epsilon^{(\phi)}_{\star}}\right)^{\frac{1}{2}},\hspace{30pt}
      \frac{\partial\phi_{f}}{\partial\chi_{\star}}=-\frac{H_{f}}{H_{\star}}\,\frac{\epsilon^{(\chi)}_{f}}{\epsilon_{f}}\,
      \left(\frac{c^{(\phi)}_{f}}{c^{(\chi)}_{\star}}\,
      \frac{\epsilon^{(\phi)}_{f}}{\epsilon^{(\chi)}_{\star}}\right)^{\frac{1}{2}},\nonumber\\[10pt]
      \frac{\partial\chi_{f}}{\partial\chi_{\star}}&=\frac{H_{f}}{H_{\star}}\,
      \frac{\epsilon^{(\phi)}_{f}}{\epsilon_{f}}\,\left(\frac{c^{(\chi)}_{f}}{c^{(\chi)}_{\star}}\,
      \frac{\epsilon^{(\chi)}_{f}}{\epsilon^{(\chi)}_{\star}}\right)^{\frac{1}{2}},\hspace{30pt}
      \frac{\partial\chi_{f}}{\partial\phi_{\star}}=-\frac{H_{f}}{H_{\star}}\,
      \frac{\epsilon^{(\phi)}_{f}}{\epsilon_{f}}\,
      \left(\frac{c^{(\chi)}_{f}}{c^{(\phi)}_{\star}}\,
      \frac{\epsilon^{(\chi)}_{f}}{\epsilon^{(\phi)}_{\star}}\right)^{\frac{1}{2}},\label{eq:field-derivatives}
      \end{align}
      \vspace{5pt}

\noindent which are used to evaluate the derivatives of $N$ in the following section.

\subsection{Results\label{sec:separable-results}}

Given the expressions (\ref{eq:dN-dphi}) and the results (\ref{eq:field-derivatives}), the first derivatives of $N$ can be calculated in terms of slow variation parameters
      
      \vspace{5pt}
      \be N_{,\phi_{\star}}=\frac{1}{\sqrt{2\epsilon^{(\phi)}_{\star}c^{(\phi)}_{\star}}}u,\hspace{30pt}N_{,\chi_{\star}}=\frac{1}{\sqrt{2\epsilon^{(\chi)}_{\star}c^{(\chi)}_{\star}}}v.\label{eq:first-derivatives}\ee

\newpage

\noindent For brevity we have introduced the following definitions 

      \vspace{-5pt}
      \be u=\frac{H^{(\phi)}_{\star}+Z_{f}}{H_{\star}},\hspace{30pt}
      v=\frac{H^{(\chi)}_{\star}-Z_{f}}{H_{\star}},\hspace{30pt}
      Z_{f}=\frac{H^{(\chi)}_{f}\epsilon^{(\phi)}_{f}-H^{(\phi)}_{f}\epsilon^{(\chi)}_{f}}{\epsilon_{f}}.\ee
			\vspace{5pt}

\noindent The above neatly demonstrates the nature of the $\delta N$ formalism, since the term $Z_{f}$ controls the super-horizon evolution whilst the remaining terms are evaluated at horizon exit. To calculate $n_{\zeta}$ and $\fnl^{(4)}$ we also require the second derivatives of $N$, which are found by simply differentiating (\ref{eq:first-derivatives}) to give
		
      \vspace{-5pt}
      \begin{align}
      N_{,\phi_{\star}\phi_{\star}}&=\frac{1}{2}\left(\frac{1}{c^{(\phi)}_{\star}}
      -\frac{\left(\eta^{(\phi)}_{\star}+s^{(\phi)}_{\star}\right)}{\epsilon^{(\phi)}_{\star}c^{(\phi)}_{\star}}u
      +\frac{\sqrt{2}}{\sqrt{\epsilon^{(\phi)}_{\star}c^{(\phi)}_{\star}}}\frac{Z_{f,\phi_{\star}}}{H_{\star}}\right),\nonumber\\[12.5pt]
      N_{,\chi_{\star}\chi_{\star}}&=\frac{1}{2}\left(\frac{1}{c^{(\chi)}_{\star}}
      -\frac{\left(\eta^{(\chi)}_{\star}+s^{(\chi)}_{\star}\right)}{\epsilon^{(\chi)}_{\star}c^{(\chi)}_{\star}}v
      -\frac{\sqrt{2}}{\sqrt{\epsilon^{(\chi)}_{\star}c^{(\chi)}_{\star}}}\frac{Z_{f,\chi_{\star}}}{H_{\star}}\right),\nonumber\\[12.5pt]
      N_{,\phi_{\star}\chi_{\star}}&=\frac{1}{\sqrt{2\epsilon^{(\phi)}_{\star}c^{(\phi)}_{\star}}}\frac{Z_{f,\chi_{\star}}}{H_{\star}}=
      -\frac{1}{\sqrt{2\epsilon^{(\chi)}_{\star}c^{(\chi)}_{\star}}}\frac{Z_{f,\phi_{\star}}}{H_{\star}},\label{eq:second-derivatives}
      \end{align}
      \vspace{10pt}
				
\noindent where the derivatives of $Z_{f}$ are given by

      \be Z_{f,\phi_{\star}}=\frac{\sqrt{2}H_{\star}}{\sqrt{c^{(\phi)}_{\star}\epsilon^{(\phi)}_{\star}}}\mathcal{A},\hspace{30pt} Z_{f,\chi_{\star}}=-\frac{\sqrt{2}H_{\star}}{\sqrt{c^{(\chi)}_{\star}\epsilon^{(\chi)}_{\star}}}\mathcal{A}.\ee
      \vspace{3pt}

\noindent Again for brevity we have introduced the following definition:
      
      \vspace{3pt}
      \be \mathcal{A}=-\frac{H_{f}^{2}}{H_{\star}^{2}}\frac{\epsilon^{(\phi)}_{f}\epsilon^{(\chi)}_{f}}{\epsilon_{f}}\left(\frac{1}{2}-\frac{\eta_{f}^{ss}}{\epsilon_{f}}-\frac{1}{2}\frac{s^{ss}_{f}}{\epsilon_{f}}\right).\label{eq:A}\ee
      \vspace{7pt}

\noindent This is equivalent to the parameter $\mathcal{A}$ in \cite{Byrnes09} and both expressions contain the second term in parenthesis
      
      \vspace{-2.5pt}
      \be \eta^{ss}=\frac{\epsilon^{(\chi)}\eta^{(\phi)}+\epsilon^{(\phi)}\eta^{(\chi)}}{\epsilon}.\ee
      \vspace{1pt}

\newpage

\noindent We find an additional term however, specific to non-standard kinetic terms, given by

      \vspace{2pt}
      \be s^{ss}=\frac{\epsilon^{(\chi)}s^{(\phi)}+\epsilon^{(\phi)}s^{(\chi)}}{\epsilon}.\label{eq:sss}\ee
      \vspace{4pt}

\noindent Finally, we substitute the results for $N_{,I}$ (\ref{eq:first-derivatives}) and $N_{,IJ}$ (\ref{eq:second-derivatives}) into the expressions for $n_{\zeta}$ (\ref{eq:spectral-index-deltaN}) and $\fnl^{(4)}$ (\ref{eq:fnl4}) to find 
  
      \vspace{5pt}
      \be n_{\zeta}-1=-2\epsilon_{\star}-2\frac{
      u\left(1-\frac{(\eta^{(\phi)}_{\star}+s^{(\phi)}_{\star})}{\epsilon^{(\phi)}_{\star}}u\right)+v\left(1-\frac{(\eta^{(\chi)}_{\star}+s^{(\chi)}_{\star})}{\epsilon^{(\chi)}_{\star}}v\right)}{\frac{u^{2}}{\epsilon^{(\phi)}_{\star}}+\frac{v^{2}}{\epsilon^{(\chi)}_{\star}}},\label{eq:spectral-index-result}\ee
    
      \vspace{5pt}

      \be \frac{6}{5}\fnl^{(4)}=\frac{\frac{u^{2}}{\epsilon^{(\phi)}_{\star}}\left(1-\frac{(\eta^{(\phi)}_{\star}+s^{(\phi)}_{\star})}
      {\epsilon^{(\phi)}_{\star}}u\right)+
      \frac{v^{2}}{\epsilon^{(\chi)}_{\star}}\left(1-\frac{(\eta^{(\chi)}_{\star}+s^{(\chi)}_{\star})}
      {\epsilon^{(\chi)}_{\star}}v\right)+
      2\left(\frac{u}{\epsilon^{(\phi)}_{\star}}-\frac{v}{\epsilon^{(\chi)}_{\star}}\right)^{2}\mathcal{A}}
      {\left(\frac{u^{2}}{\epsilon^{(\phi)}_{\star}}+\frac{v^{2}}{\epsilon^{(\chi)}_{\star}}\right)^{2}}. 
     \label{eq:fnl4-result}\ee
      \vspace{5pt}

\noindent The above results are valid for the two-field scenario detailed in section~\ref{sec:model-homogeneous} assuming comparable sound speeds and slow variation at horizon exit, in addition to a separable Hubble parameter. Note that $c^{(\phi)}_{\star}$ and $c^{(\chi)}_{\star}$ do not appear explicitly in these expressions, since we assume they are approximately equal and therefore cancel.   

It is straight forward to confirm that (\ref{eq:spectral-index-result}) and (\ref{eq:fnl4-result}) recover previous results in the appropriate limits. To recover the single field expressions we consider $\dot{\chi}\rightarrow0$, such that $Z_{f}\rightarrow H_{f}^{(\chi)}$, $u\rightarrow1$ and $v\rightarrow0$, since in the single field limit the curvature perturbation $\zeta$ is conserved after horizon exit and all parameters can be evaluated at $t_{\star}$. With this, we find
	
      \vspace{3pt}
      \be n_{\zeta}-1=-4\epsilon_{\star}^{(\phi)}+2\eta_{\star}^{(\phi)}+2s_{\star}^{(\phi)},\hspace{30pt}\frac{6}{5}\fnl^{(4)}=\epsilon_{\star}^{(\phi)}-\eta_{\star}^{(\phi)}-s_{\star}^{(\phi)},\ee
      \vspace{3pt}

\noindent and recover the single field results \cite{Chen07}. Moreover, it is trivial to check that we recover the canonical results, since in the limit $c^{(\phi)}\hspace{-3pt}=c^{(\chi)}\hspace{-3pt}=1$ the expressions (\ref{eq:spectral-index-result}) and (\ref{eq:fnl4-result}) coincide with those of \cite{Byrnes09}.

With regard to the general behaviour of our results, inspection of (\ref{eq:spectral-index-result}) shows that $n_{\zeta}-1$ will remain $\mathcal{O}(\epsilon_{\star})$, where here $\epsilon$ represents first order in slow variation. This is in keeping with observation of a nearly scale invariant power spectrum \cite{Komatsu10}. Notice that whilst the above does contain additional factors of $s^{(\phi)}_{\star}$ and $s^{(\chi)}_{\star}$ with respect to \cite{Byrnes09}, their effect does not alter the conclusion that $n_{\zeta}-1$ remains $\mathcal{O}(\epsilon_{\star})$.

Turning to $\fnl^{(4)}$, the first two terms in parenthesis in (\ref{eq:fnl4-result}) are similarly $\mathcal{O}(\epsilon_{\star})$ and so would contribute little towards any enhanced non-Gaussianity. Moreover, barring certain fine-tuned choices for the trajectory, the prefactor of $\mathcal{A}$ is generally $\mathcal{O}(1)$ (see \cite{Byrnes08} for a summary of when this term is large in the canonical case). As a result, $\mathcal{A}$ provides the only potential source of significant non-Gaussianity. Whilst the terms preceding the parenthesis in the expression for $\mathcal{A}$ (\ref{eq:A}) are $\mathcal{O}(\epsilon_{\star})$, $\eta^{ss}_{f}$ and $s^{ss}_{f}$ can become much larger than unity. In the canonical case for example, \cite{Byrnes09} show that $\eta^{ss}_{f}$ can produce enhanced non-Gaussianity as the trajectory descends a potential in which $\eta^{(\phi)}$ and $\eta^{(\chi)}$ become large. For non-trivial sound speeds however, we find the additional factor $s^{ss}_{f}$ which can, in principle, produce a similar contribution. For example, one might expect an analogous scenario to \cite{Byrnes09} in which $s^{ss}_{f}$ produces significant non-Gaussianity as the trajectory traverses a warped region in which $c^{(\phi)}$ and $c^{(\chi)}$ change abruptly and so $s^{(\phi)}$ and $s^{(\chi)}$ are enhanced. In the following section we will apply our general formulae to such a scenario and demonstrate how $s^{ss}_{f}$ becomes large, producing enhanced local non-Gaussianity during inflation.

We note here that the expressions (\ref{eq:spectral-index-result}) and (\ref{eq:fnl4-result}) are well suited for describing the evolution of $n_{\zeta}$ and $\fnl^{(4)}$ \emph{during} inflation. However, these are not necessarily the observed values since in general we must track their evolution from the end of inflation until they are imprinted on the cosmic microwave background (CMB) at decoupling. Ideally, one would know the entire evolution history between these times but, given our lack of knowledge of the early Universe, this is not currently feasible. In light of this, recent work has considered whether such non-Gaussianity produced during inflation can imprint upon the CMB \cite{Elliston11a,Elliston11b,Meyers11-a,Meyers11-b,Peterson11a,Watanabe11,Choi12,Mazumdar12}. For example, \cite{Elliston11a,Elliston11b} consider scenarios in which a limit is reached during or shortly after inflation, in which perturbations become essentially adiabatic and approximately conserved. It may be that the potential has a focusing region in which trajectories converge before the end of inflation and, in certain circumstances, can imprint upon the CMB. Alternatively, a waterfall field may end inflation whilst $\fnl^{(4)}$ is large, after which it is preserved. If a conserved limit is reached shortly after inflation however, one must still choose a method for reheating and the results may depend upon this choice. As a further example, \cite{Meyers11-a,Meyers11-b} consider sum separable potentials and find non-Gaussianity is damped during the transition to single field behaviour, leaving little imprint on the CMB. Such study is invaluable for understanding what we can infer about inflationary dynamics from observations of non-Gaussianity and it would be interesting to make similar considerations in our case. In this work however, we simply illustrate the potential production of non-Gaussianity through multiple-field dynamics during inflation, facilitated by non-trivial sound speeds as opposed to the potential.      

\section{Inflation in two cutoff throats\label{sec:example}}

To illustrate the more general arguments presented in the previous section, we present a specific scenario that can be described by the above framework. In particular, we consider two-field inflation in the tip regions of two warped throats. We find that the combined effect of rapidly varying sound speeds and multiple-field dynamics can lead to significant local non-Gaussianity towards the end of inflation, for certain initial conditions.

\newpage

      \begin{figure}[t]
      \begin{center}
      \includegraphics[scale=0.625]{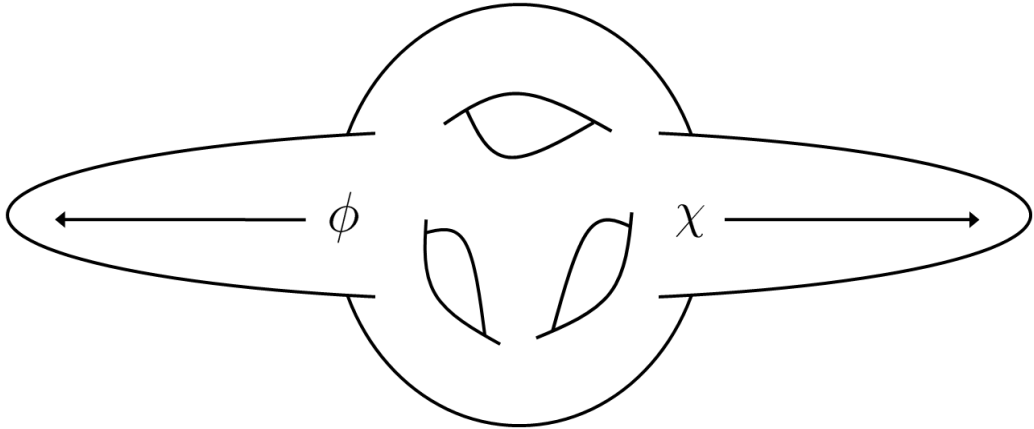}		
      \caption{Geometric illustration of our example model. Inflation is driven by two probe D3 branes traversing two distinct warped throats glued to a compact Calabi-Yau manifold in type IIB string theory. The radial co-ordinates of the D-branes play the role of the fields $\phi$ and $\chi$. The throats are produced by stacks of D3 branes that source RR fluxes and we include fractional D3 branes at the tips to cut-off the throats.\label{fig:diagram}}
      \end{center}
      \end{figure} 

\subsection{Background trajectory\label{sec:example-trajectory}}

We model inflation as driven by two probe D3 branes traversing two distinct warped throats glued to a compact Calabi-Yau in type IIB string theory \cite{Cai09,Cai09-1,Pi11}, as in figure~\ref{fig:diagram}. Each throat is produced by a stack of N D3 branes located at the tip of the conifold, sourcing N units of 5-form flux which warp the space. In addition, we add M wrapped D5 branes at the apex of each throat that source M units of magnetic RR 3-form flux, through the $S^{3}$ of the basis $T^{11}$ of the cone. The resulting space-time is a Klebanov-Tseytlin throat \cite{Klebanov00-a} and we refer to \cite{Herzog02} for a review of the geometrical aspects of this set up and \cite{Kecskemeti06} for the case of single field inflation in a cut-off throat. Given the above, the warp factors are given by

\begin{figure}[t]
      \begin{center}
      \includegraphics[scale=0.43]{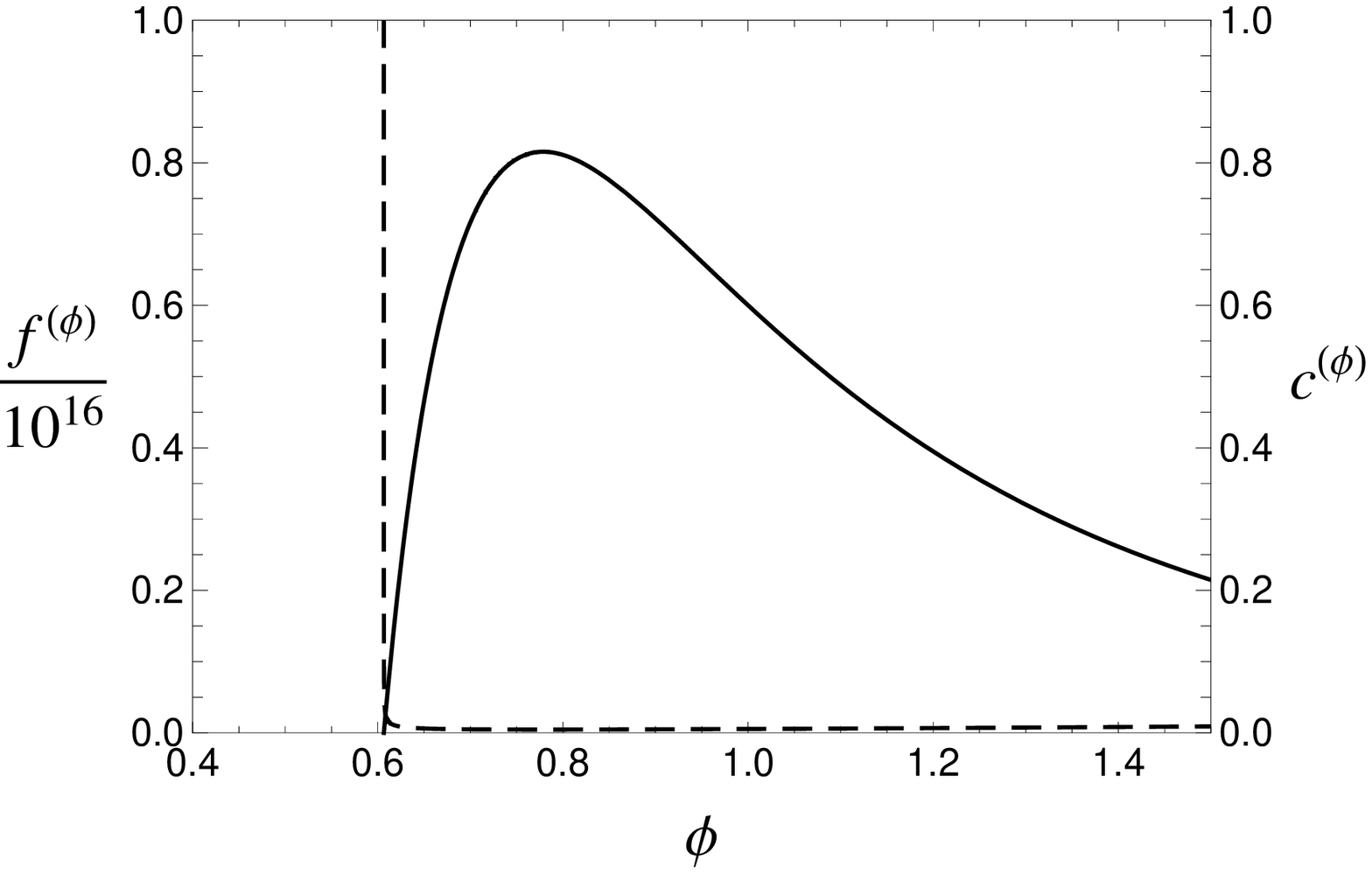}
      \includegraphics[scale=0.43]{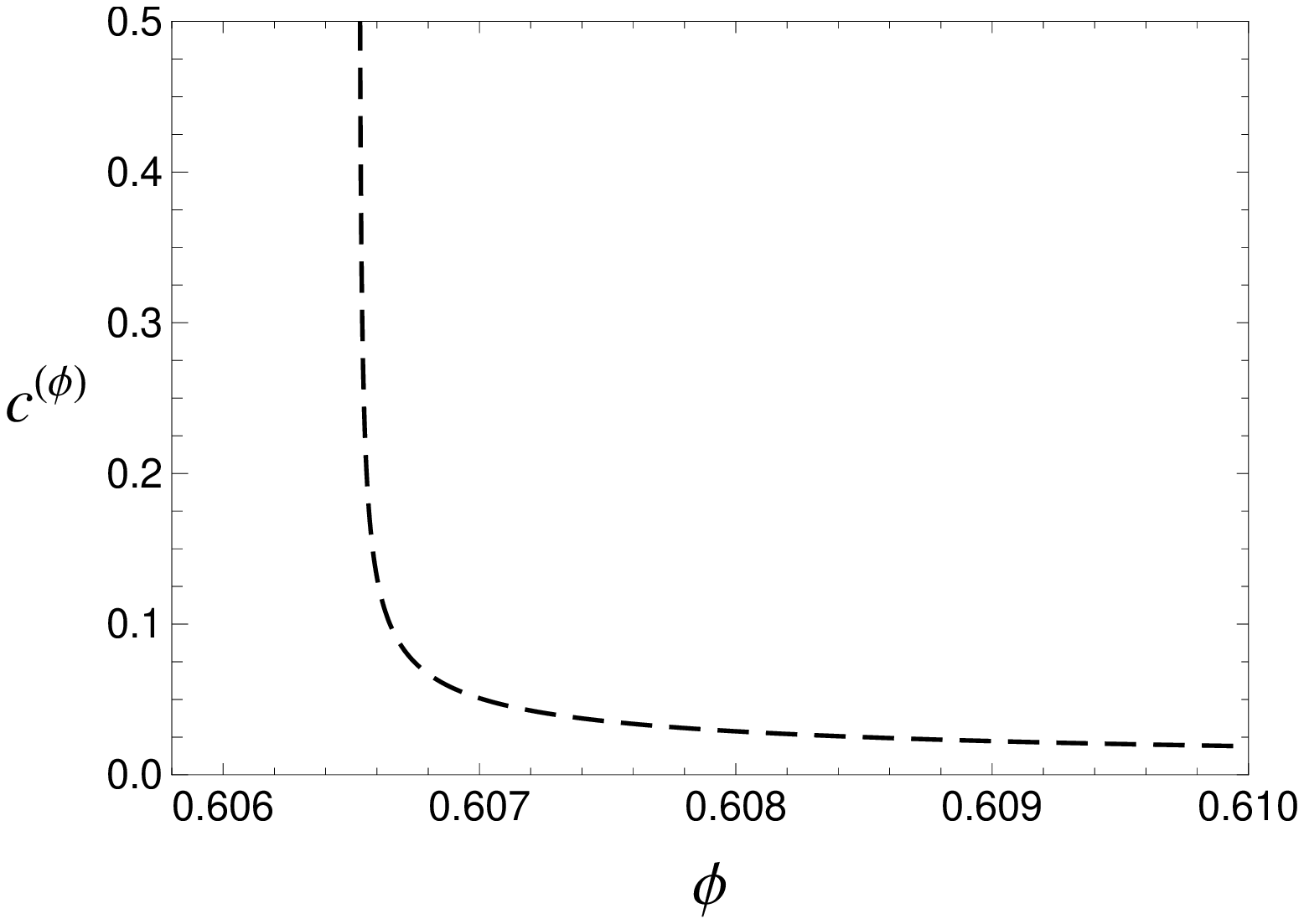}
      \caption{\emph{Left:} The warp factor $f^{(\phi)}$ (solid) and corresponding sound speed $c^{(\phi)}$ (dashed) plotted as functions of $\phi$, with $\lambda_{1}=6\times10^{15}$, $\lambda_{2}=2$, $\lambda_{3}=1$ and $\mathcal{H}^{(\phi)}=1.188\times10^{-6}$. \emph{Right:} The rapid increase in $c^{(\phi)}$ towards the tip of the throat in further detail.\label{fig:warpfactor-soundspeed}}
      \end{center}
      \end{figure}

      \vspace{-1pt}
      \be f^{(\phi)}=\frac{\lambda_{1}}{\phi^{4}}\left(1+\lambda_{2}\,\log\left(\frac{\phi}{\lambda_{3}}\right)\right),\label{eq:warp-factor}\ee
      \vspace{5pt}

\noindent where $f^{(\chi)}$ is given by replacing $\phi\rightarrow\chi$. For simplicity we have assumed the same warping in each throat, such that $\lambda^{(\phi)}_{i}=\lambda^{(\chi)}_{i}=\lambda_{i}$, which does not qualitatively alter our conclusions. Here $\lambda_{i}$  control the number of RR fluxes switched on and, in particular, $\lambda_{2}$ is dependent on  $M$ \cite{Herzog02}. Geometrically, the logarithmic running of the warp factor implies that the effective number of units of 5-form flux varies, depending on the position along the throat. Note the infrared singularity at $\phi=\lambda_{3}\,e^{-1/\lambda_{2}}$ where the supergravity approximation used to obtain the geometry breaks down and the  solution becomes unreliable. By considering a generalization of the previous set-up, obtained by wrapping  branes on a deformed conifold, this singularity can be tamed \cite{Klebanov00-b}. However, the inflationary trajectory we consider does not reach the IR singularity and so we adopt (\ref{eq:warp-factor}) for the warp factors. Figure~\ref{fig:warpfactor-soundspeed} illustrates $f^{(\phi)}$ for representative values of the parameters, which we will use hereafter. Note that such choices are used for illustrative purposes only, since a full exploration of parameter space is beyond the scope of this work. Given this, we choose a linear, separable Hubble parameter

      \vspace{-3pt}
      \be H(\phi,\chi)=\mathcal{H}^{(\phi)}\phi+\mathcal{H}^{(\chi)}\chi,\ee

      \vspace{0pt}

\noindent such that $\eta^{(\phi)}\hspace{-3pt}=\eta^{(\chi)}\hspace{-3pt}=0$. This guarantees that the only significant contribution towards $\fnl^{(4)}$ in (\ref{eq:fnl4-result}) must be $s^{ss}$, since $\eta^{ss}=0$ throughout inflation. Finally, with expressions for the Hubble parameter and warp factors, the potential is fixed by the Friedmann equation (\ref{eq:friedman-dbi}) and, throughout most of inflation, is well approximated by 
		
      \vspace{-3pt}
      \be V(\phi,\chi)\simeq3\mathcal{H}^{(\phi)^{2}}\phi^{2}+3\mathcal{H}^{(\chi)^{2}}\chi^{2}+6\mathcal{H}^{(\phi)}\mathcal{H}^{(\chi)}\phi\chi.\ee \vspace{-4pt}

\noindent Whilst this potential is required to suit our analytical approach, it allows us to demonstrate properties that we expect to be representative of a broader class of potentials. Using (\ref{eq:sound-speed-dbi}), we arrive at the following expression for the sound speeds

      \vspace{3pt}
      \be c^{(\phi)}=\frac{1}{\sqrt{1+\frac{4\mathcal{H}^{(\phi)^{2}}\lambda_{1}}{\phi^{4}}\left(1+\lambda_{2}\,\log\left(\frac{\phi}{\lambda_{3}}\right)\right)}},\ee
      \vspace{6pt}

\noindent where $c^{(\chi)}$ is again given by replacing $\phi\rightarrow\chi$. Figure~\ref{fig:warpfactor-soundspeed} illustrates $c^{(\phi)}$ for the $f^{(\phi)}$ plotted previously. This provides some qualitative insight into the inflationary dynamics. Considering for a moment the single field case, inflation progresses efficiently with $c^{(\phi)}\ll1$ and $\epsilon^{(\phi)}\ll1$ for $\phi\gtrsim0.7$, as in the standard single field DBI case. For $\phi\lesssim0.7$ however, corresponding to the tip of the throat, the sound speed rises rapidly and drives $\epsilon^{(\phi)}$ towards unity, where inflation ends. Similar qualitative arguments hold for the two-field case but the evolution will depend more on the choice of model parameters.

\begin{figure}[t]
      \begin{center}
      \includegraphics[scale=0.5]{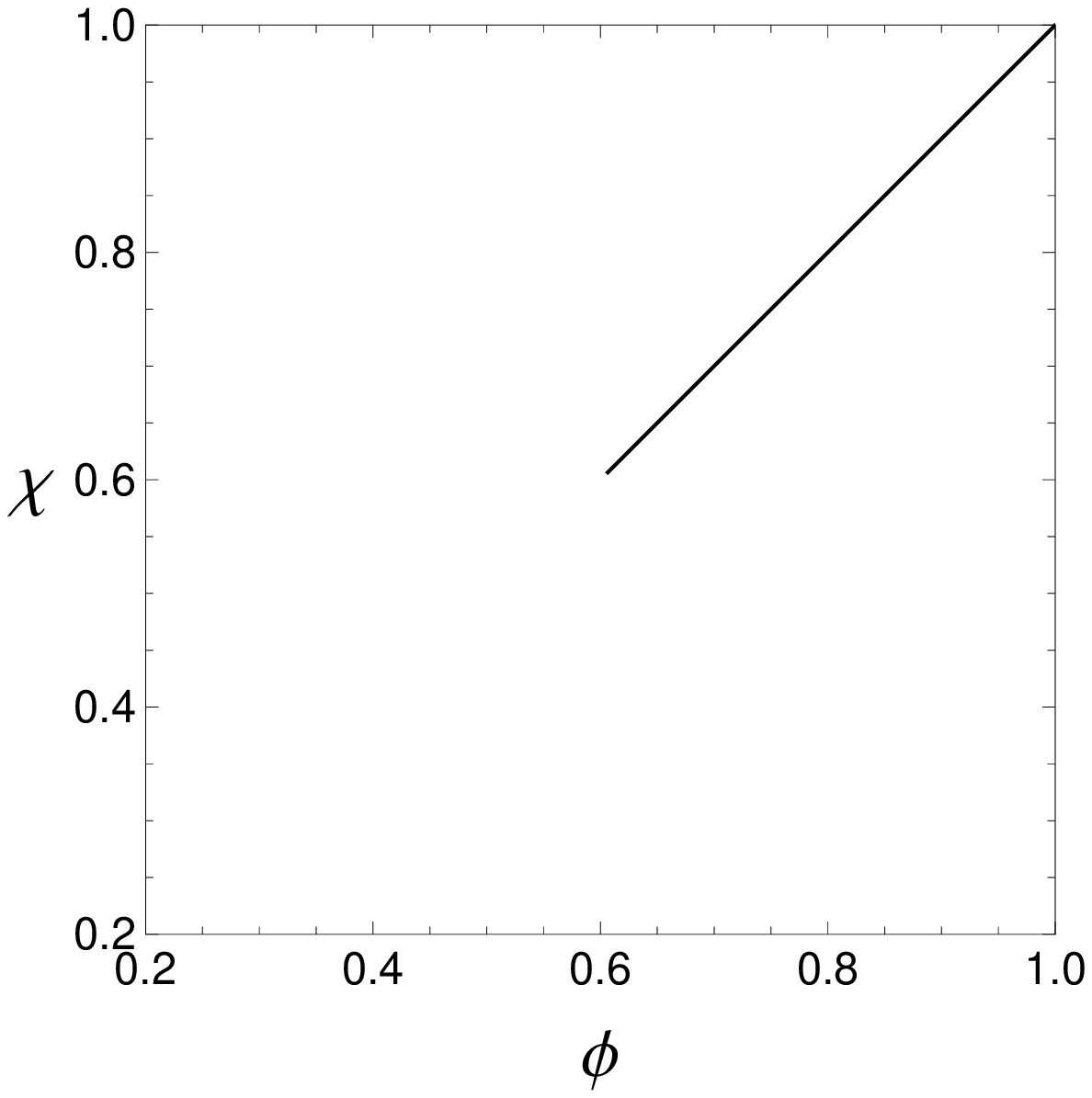}
      \includegraphics[scale=0.5]{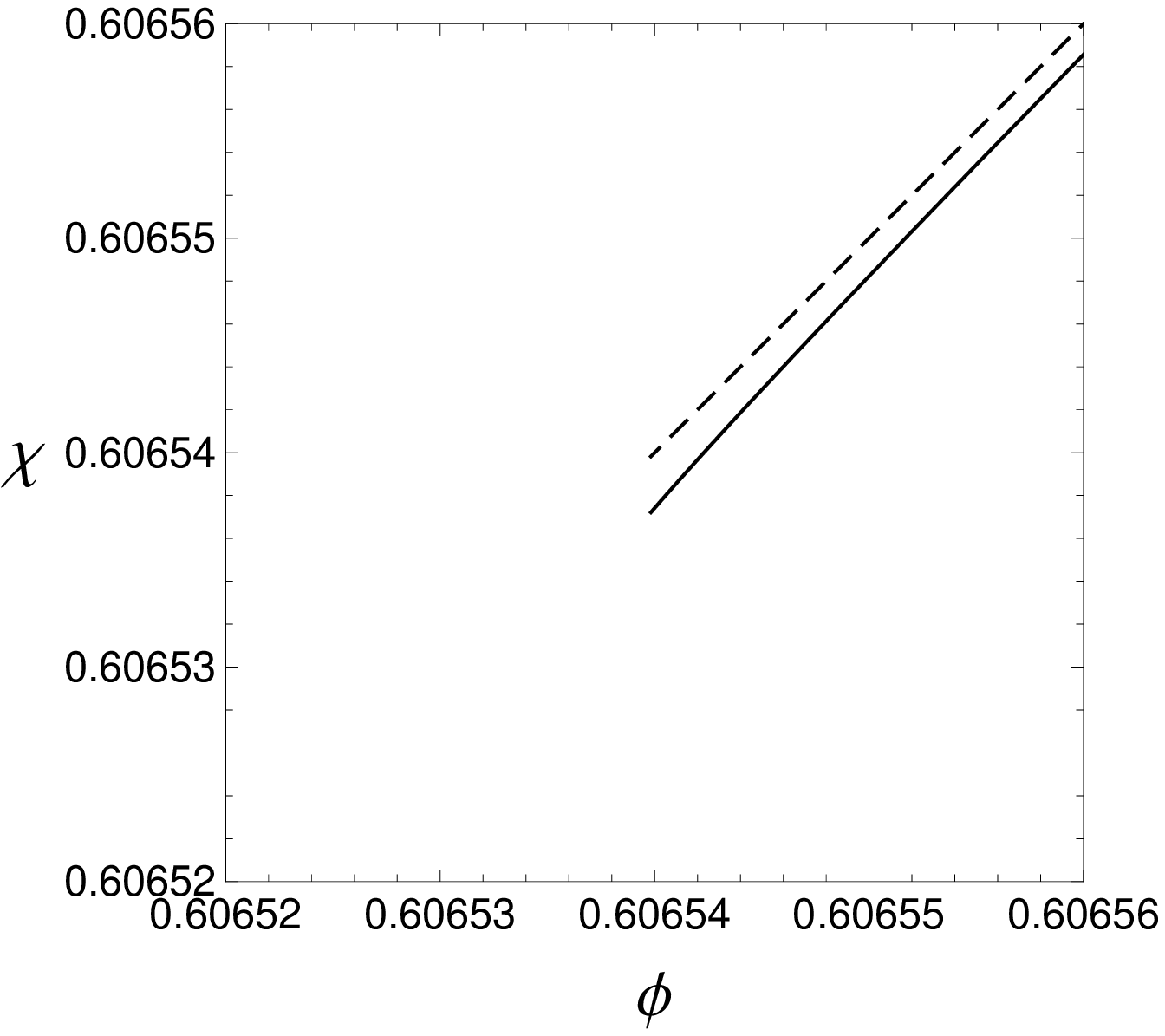}
      \caption{\emph{Left:} The trajectory in field space, originating at $\phi(t_{\star})=\chi(t_{\star})=1$ and ending after $N\simeq60$ e-folds of inflation, when $\epsilon=1$. \emph{Right:} Enlarged region of the trajectory (solid) illustrating the turn in the $\chi$ direction towards the end of inflation. The dashed line shows the straight line trajectory corresponding to $\mathcal{H}^{(\phi)}=\mathcal{H}^{(\chi)}$ for comparison.\label{fig:field-trajectory}}
      \end{center}
      \end{figure}

To study the precise dynamics we must solve the equations of motion. Since the Friedmann equation is automatically satisfied we need only solve the field equations (\ref{eq:field-dbi}) for a given choice of parameters and initial conditions. As such we choose $\lambda_{1}=6\times10^{16}$, which is typically required in standard DBI \cite{Alishahiha04}, $\lambda_{2}=2$ and $\lambda_{3}=1$ to characterise the warp factors. Furthermore, we choose $\phi(t_{\star})=\chi(t_{\star})=1$ as our initial conditions, where $t_{\star}$  is  the time at which observable scales exit the horizon. We require $N\simeq60$ e-folds between $t_{\star}$ and the end of inflation, which we choose to define as $\epsilon=1$. Note that this choice is a working definition for the end of inflation, since we do not consider an explicit model of reheating in this work. The choice of parameters and initial conditions thus far is `symmetric', in the sense that for $\mathcal{H}^{(\phi)}=\mathcal{H}^{(\chi)}$ the resultant trajectory would be a straight line. To facilitate multiple-field behaviour we introduce a small asymmetry such that $\mathcal{H}^{(\phi)}=1.188\times10^{-6}$ and $\mathcal{H}^{(\chi)}=1.192\times10^{-6}$.  With this choice of parameters $c^{(\phi)}/c^{(\chi)}\sim1+10^{-3}$ at $t_{\star}$ and is therefore consistent with our approximation that the sound horizons are comparable when observable scales exit. Given this, we solve the field equations and plot the inflationary trajectory in figure~\ref{fig:field-trajectory}. 

\begin{figure}[t]
      \begin{center}
      \includegraphics[scale=0.53]{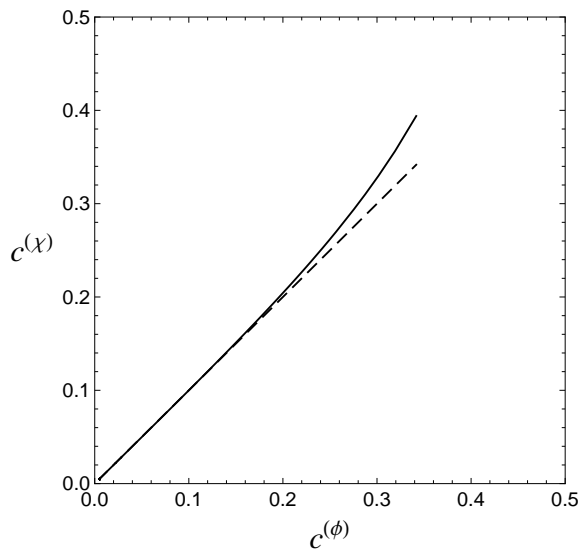}
      \caption{Evolution of the sound speeds $c^{(\phi)}$ and $c^{(\chi)}$ (solid). For most of inflation $c^{(\phi)}\simeq c^{(\chi)}\ll1$ but during the last few e-folds both rise rapidly, with $c^{(\chi)}>c^{(\phi)}$. This explains the downward curving trajectory in figure~\ref{fig:field-trajectory}. For comparison, the dashed line shows the corresponding evolution for $\mathcal{H}^{(\phi)}=\mathcal{H}^{(\chi)}$.  \label{fig:sound-speed-trajectory}}
      \end{center}
      \end{figure}

As we would expect, given the almost symmetric initial conditions and parameter choices, the trajectory is approximately straight. However, towards the end of inflation the trajectory curves in the $\chi$ direction. To understand this, we show the evolution of the sound speeds parametrically in figure~\ref{fig:sound-speed-trajectory}. Throughout most of inflation $c^{(\phi)}\simeq c^{(\chi)}\ll1$, behaving essentially as standard single field DBI. At the tip of the throats, corresponding to the last few e-folds of inflation, both sound speeds increase rapidly. By choosing $\mathcal{H}^{(\chi)}>\mathcal{H}^{(\phi)}$ however, $c^{(\chi)}$ increases slightly before $c^{(\phi)}$, producing the upwards turn in figure~\ref{fig:sound-speed-trajectory}. \noindent Since the motion of $\chi$ is now less inhibited than that of $\phi$, the trajectory in figure~\ref{fig:field-trajectory} curves in the $\chi$ direction. Whilst this implies the presence of isocurvature modes at the end of inflation, we cannot track any subsequent evolution without an explicit model of reheating, as noted in section~\ref{sec:separable-results}.

\subsection{Evolution of $\fnl^{(4)}$\label{sec:example-observables}}

Given this trajectory, we use the expressions in section~\ref{sec:separable-results} to study the evolution of the relevant quantities (e.g. $\fnl^{(4)}$) as a function of $t_{f}$ for a fixed $t_{\star}$. Considering first the two-point statistics we find $\mathcal{P}_{\zeta}=2.44\times10^{-9}$ and $n_{\zeta}-1=-0.0108$ at the end of inflation, which are consistent with observations \cite{Komatsu10} and remain approximately constant throughout inflation. This is as expected, since the corresponding expressions are not sensitive to the rapidly varying sound speeds at the end of inflation.

\begin{figure}[t]
      \begin{center}
      \includegraphics[scale=0.43]{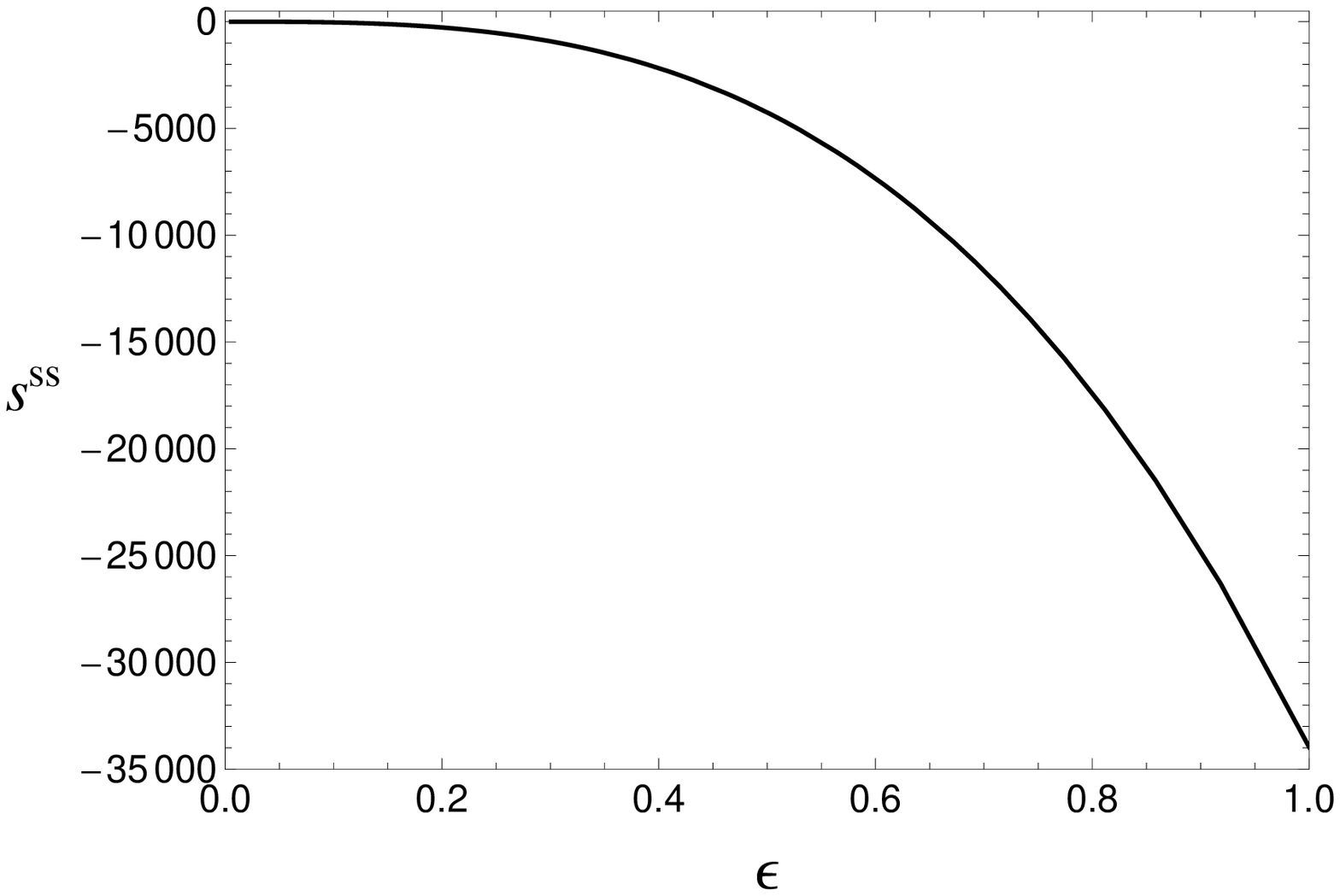}
      \includegraphics[scale=0.43]{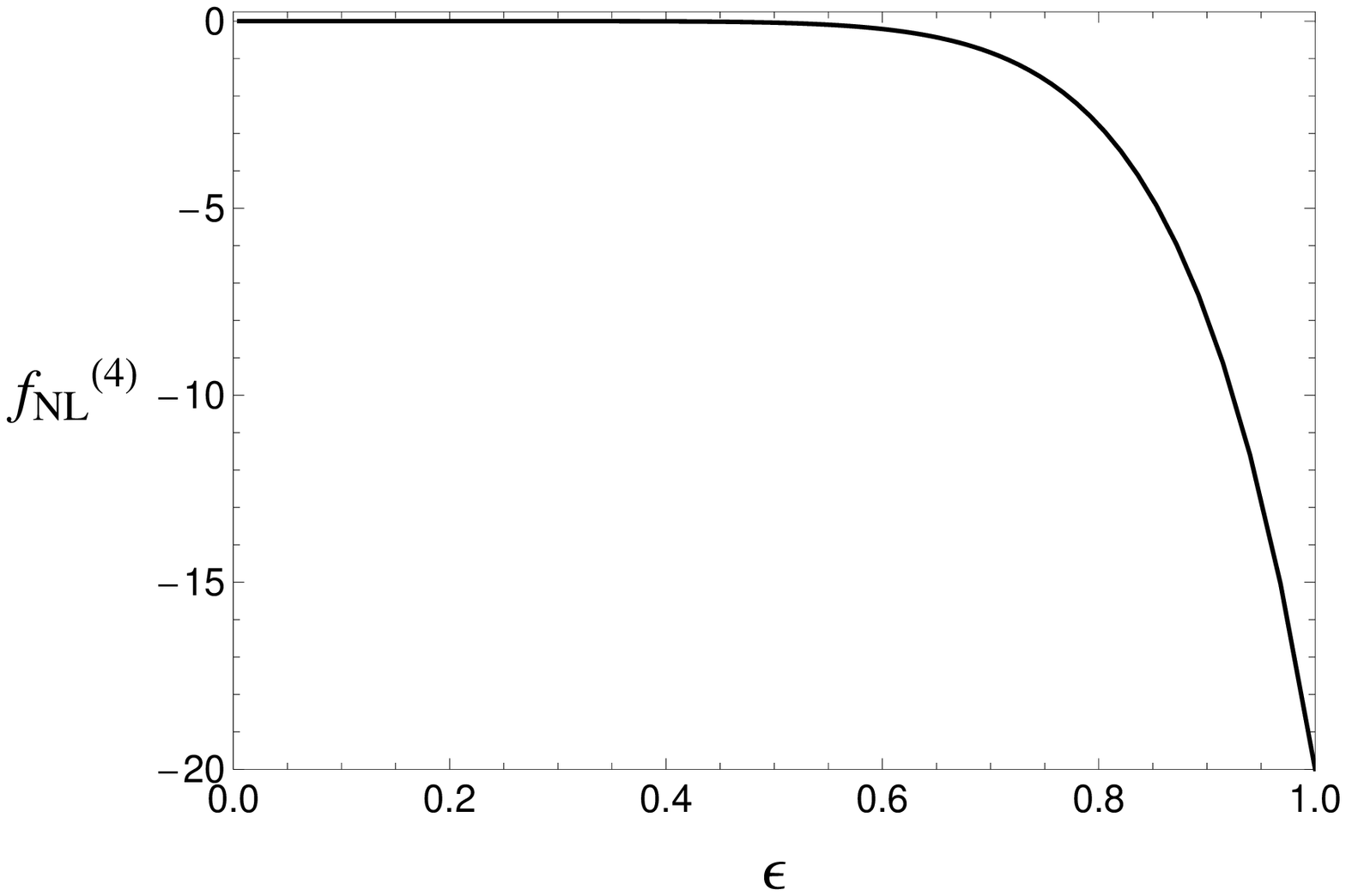}
      \caption{\emph{Left:} Evolution of $s^{ss}$ with respect to $t_{f}$, plotted as a function of $\epsilon$ for the trajectory in figure~\ref{fig:field-trajectory}. The rapidly varying sound speeds produce $s^{ss}\simeq -34000$ at the end of inflation, when $\epsilon=1$. \emph{Right:} The corresponding evolution of $\fnl^{(4)}$, where the combination of rapidly varying sound speeds and a turn in the trajectory produces $\fnl^{(4)}\simeq -20$ at the end of inflation.\label{fig:sss-fnl}}
      \end{center}
      \end{figure}

Considering now the three-point statistics, we find that the rapidly changing sound speeds allow $s^{(\phi)}$ and $s^{(\chi)}$ to increase towards the end of inflation. We note again that, since our approach does not rely on slow variation after horizon exit, we are able to study such regimes with confidence in our results. We therefore find that $s^{ss}$ increases dramatically, as illustrated in figure~\ref{fig:sss-fnl}, and so therefore does $\mathcal{A}$. In addition to the enhancement of $\mathcal{A}$, the small departure from single field behaviour towards the end of inflation, as illustrated in figure~\ref{fig:field-trajectory}, produces a non-zero term preceding $\mathcal{A}$ in the expression for $\fnl^{(4)}$ (\ref{eq:fnl4-result}). Note that regardless of the magnitude of $\mathcal{A}$, $\fnl^{(4)}$ would remain small for a straight trajectory (i.e. for $\mathcal{H}^{(\phi)}=\mathcal{H}^{(\chi)}$). The combined effect therefore is a rapid increase towards $\fnl^{(4)}\simeq-20$ at the end of inflation, as shown in figure~\ref{fig:sss-fnl}. Note that the negative sign of the non-linearity parameter in this example is the result of rapidly \emph{increasing} sound speeds, which is disfavoured but not excluded by observations \cite{Komatsu10}. It would be interesting to apply similar techniques to \cite{Elliston11a,Elliston11b} to further understand the sign of $\fnl^{(4)}$ in such scenarios. 

It is important to note that DBI type models are usually considered in the context of equilateral type non-Gaussianity, through enhanced field interactions at horizon exit. For example, for single field DBI in the equilateral configuration one finds $|\fnl^{(3)}|\sim c_{s}^{-2}$ at horizon exit, which is preserved thereafter \cite{Alishahiha04}. Moreover, $\fnl\simeq\fnl^{(3)}$ to a very good approximation, since the local contribution is negligible in the single field case. This places a firm lower bound on the sound speed at horizon exit. In the scenario considered above the sound speeds at horizon exit are similarly small and the naive expectation would be for a prohibitively large $\fnl^{(3)}$, in addition to the large $\fnl^{(4)}$ we have already discussed. The introduction of multiple field dynamics can alter this conclusion however, via conversion between entropy and adiabatic modes both during horizon exit and thereafter (see, for example, \cite{Langlois08,Langlois08b,Arroja08,RenauxPetel09} for the case of multiple field DBI in the context of a single throat, in which $\fnl^{(3)}$ is suppressed by such dynamics). We leave an explicit calculation of the equilateral contribution and its correlation with the local contribution to a follow-up paper and, for now, emphasise our main result of large local type non-Gaussianity from dynamics associated with non-trivial sound speeds. 
     
Finally we note that, given our working definition of $\epsilon=1$ as the end of inflation, at which time we evaluate the relevant quantities, we are unable to track their evolution from the end of inflation until they imprint upon the CMB. Whilst the implementation of an explicit model for reheating or the imposition of a transition to adiabaticity is outside the scope of this work, it would be an interesting topic for future research.

\subsection{The trispectrum\label{sec:example-trispectrum}}

      \begin{figure}[t]
      \begin{center}
      \includegraphics[scale=0.43]{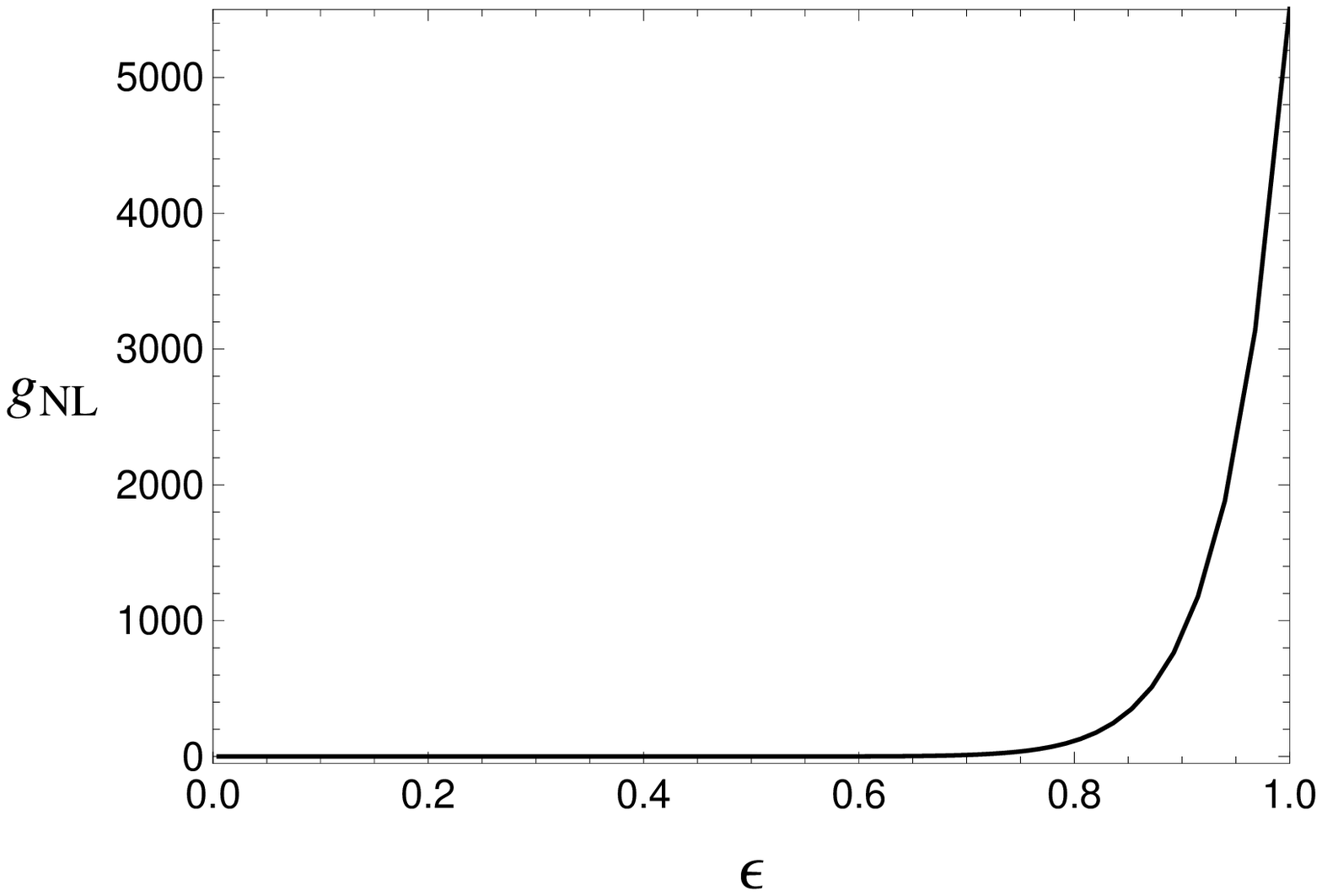}		
      \includegraphics[scale=0.43]{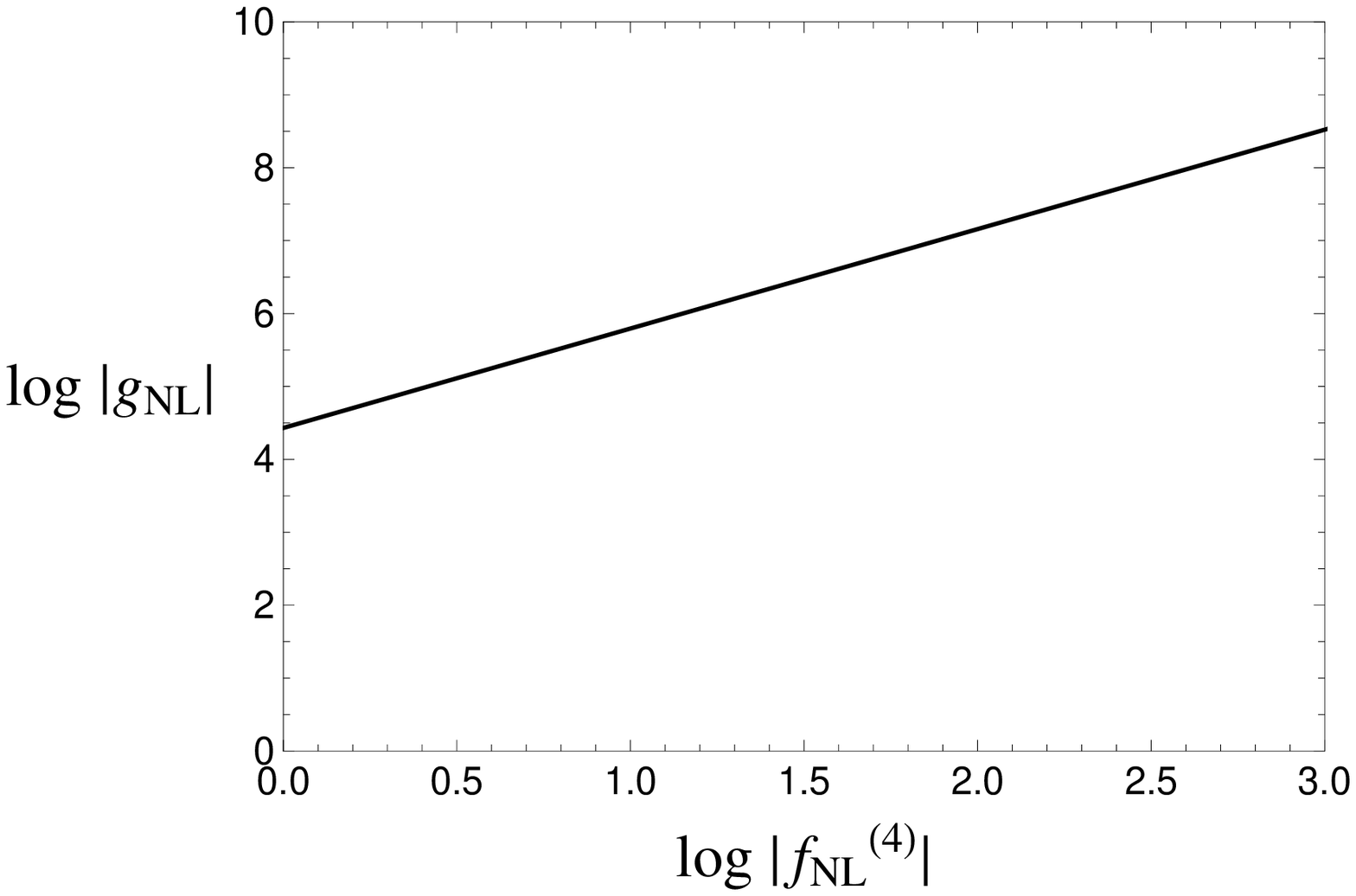}
      \caption{\emph{Left:} Evolution of $\gnl$ with respect to $t_{f}$, plotted as a function of $\epsilon$ for the trajectory in figure~\ref{fig:field-trajectory}. \emph{Right:} $\log\,|\gnl|$ as a function of $\log\,|\fnl|$. The result is a straight line with slope $\simeq 2$, such that $\gnl \propto \fnl^2$ in this case.\label{fig:gnl-loggnl}}
      \end{center}
      \end{figure} 
		
      \begin{figure}[h]
      \begin{center}
      \includegraphics[scale=0.43]{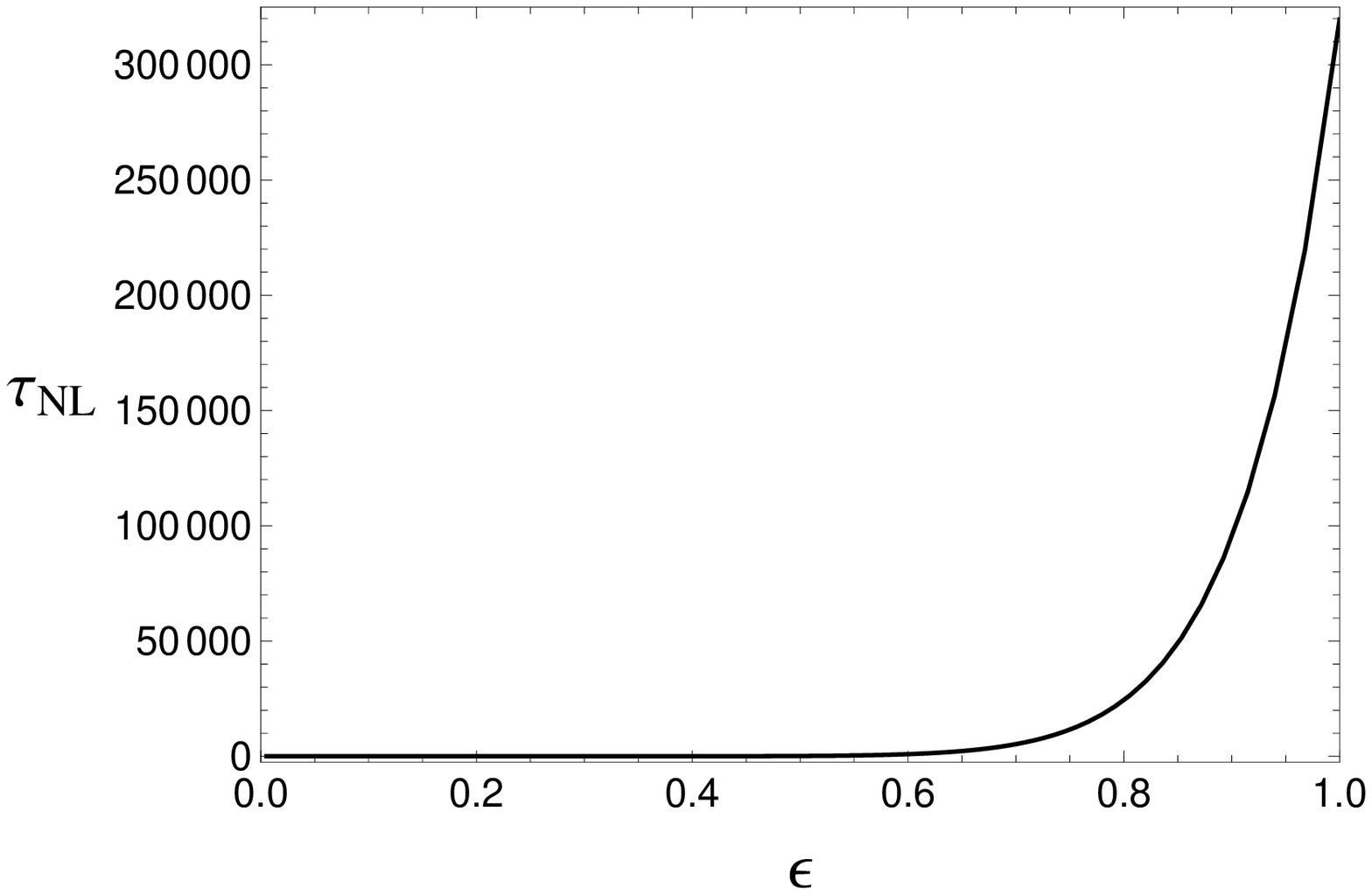}		
      \includegraphics[scale=0.43]{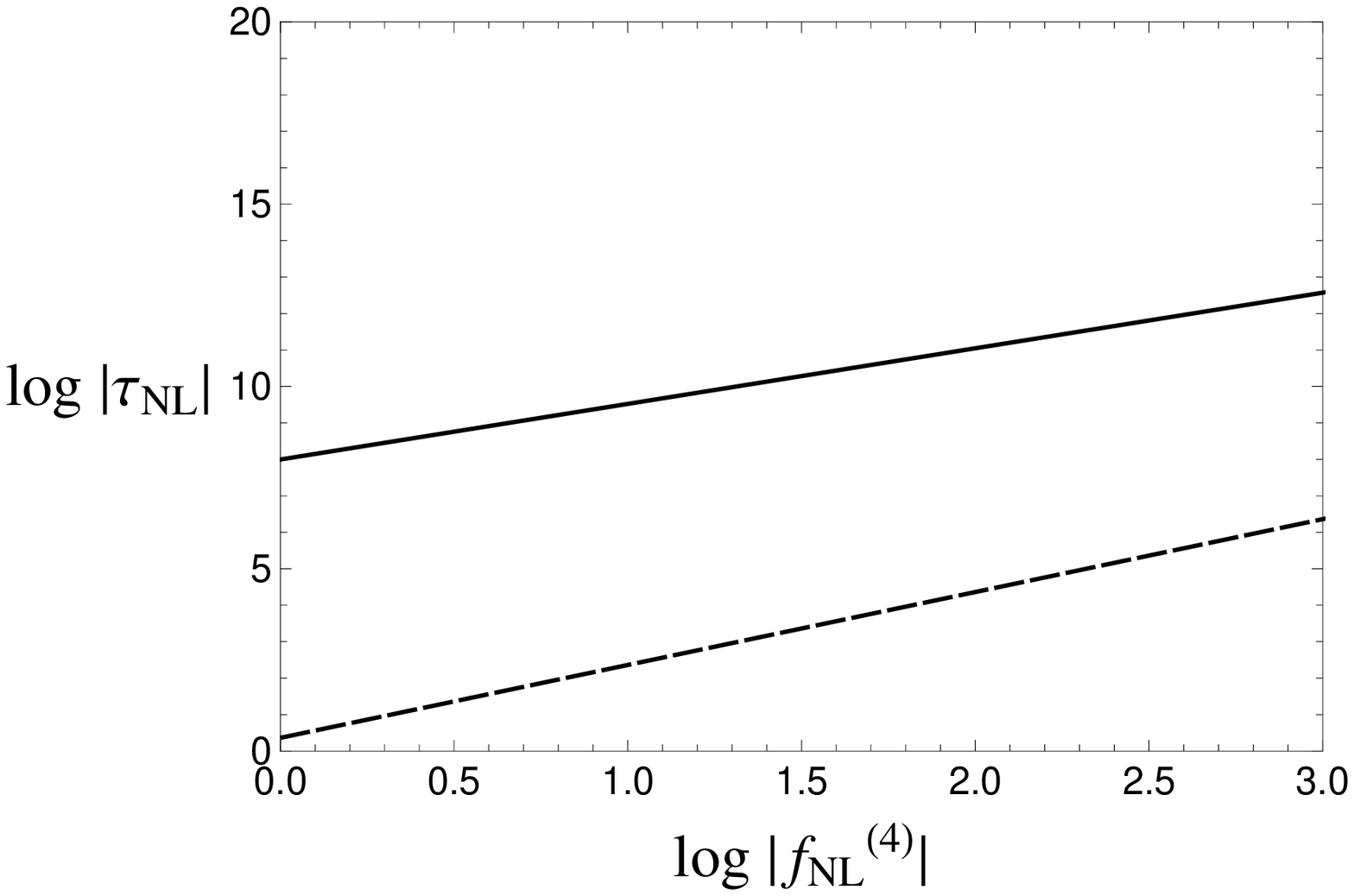}
      \caption{\emph{Left:} Evolution of $\taunl$ with respect to $t_{f}$, plotted as a function of $\epsilon$ for the trajectory in figure~\ref{fig:field-trajectory}. \emph{Right:} $\log\,|\taunl|$ as a function of $\log\,|\fnl|$ (solid). The result is a straight line with slope $\simeq 1.7$, such that $\taunl \propto \fnl^{1.7}$ in this case. The Suyama-Yamaguchi inequality (dashed) is also shown.\label{fig:taunl-logtaunl}}
      \end{center}
      \end{figure} 

In addition to the bispectrum, it is interesting to briefly consider the behaviour of the Fourier transform of the four-point function, the trispectrum. In particular, we are interested in the local-type $\gnl$ and $\taunl$, the analogues of $\fnl^{(4)}$ for the trispectrum. By performing an analogous calculation to that in section~\ref{sec:model-deltaN}, $\gnl$ and $\taunl$ can be recast using the $\delta N$ formalism \cite{Byrnes06,Alabidi06,Seery07} 

\be \gnl=\frac{25}{54}\frac{\sum_{IJK}N_{,IJK}N_{,I}N_{,J}N_{,K}}{\left(\sum_{I}N^{2}_{I}\right)^{3}},\hspace{25pt}\taunl=\frac{\sum_{IJK}N_{,IJ}N_{,IK}N_{,J}N_{,K}}{\left(\sum_{I}N^{2}_{I}\right)^{3}}.\ee
\vspace{10pt}
 
\noindent It is simple to evaluate $\taunl$ in principle, since we already have expressions for the first and second derivatives of $N$. Furthermore, we need only take a further derivative of (\ref{eq:second-derivatives}) to find the required terms in $\gnl$. In practice however, such expressions are prohibitively long. As such, we restrict ourselves to presenting their evolution along the inflationary trajectory. In figure~\ref{fig:gnl-loggnl}, $\gnl\propto\fnl^2$ and becomes $\mathcal{O}(10^{3})$ towards the end of inflation. Similarly, $\taunl$ becomes $\mathcal{O}(10^{5})$ in figure~\ref{fig:taunl-logtaunl}, where $\taunl \propto \fnl^{1.4}$. Note that the Suyama-Yamaguchi inequality \cite{Suyama08,Sugiyama11}, $\taunl\geq\left(\frac{6}{5}\fnl^{(4)}\right)^{2}$, remains satisfied by a large constant of proportionality.

\section{Conclusions\label{sec:conclusions}} 

We have studied the effect of non-trivial sound speeds and multiple-field dynamics on non-Gaussianity during inflation. In particular, we have shown that  rapid changes in the sound speeds can, in principle, produce local-type non-Gaussianity during a turn in the trajectory.

As an example of multi-component inflation with non-trivial sound speeds, we considered a multiple-DBI model and used the $\delta N$ formalism to track the super-horizon evolution of perturbations. Having expressed the homogeneous equations of motion in Hamilton-Jacobi form, we adopted the method of \cite{Byrnes09} and used a sum separable Hubble parameter to derive analytic expressions for the relevant quantities in the two-field case, valid beyond slow variation. 

We find that non-trivial sound speeds can, in principle, produce  significant local-type non-Gaussianity. Deviations from slow variation, such as rapidly varying sound speeds, enhance this effect. Note that this contribution is distinct from that of the potential and the general result is a combination of both sources. To illustrate our results we considered inflation in the tip regions of two warped throats, in which the sound speeds rapidly increase at the end of inflation. This, alongside a turn in the trajectory, produces large local-type non-Gaussianity towards the end of inflation. The non-linearity parameter is negative in this example however, owing to the rapidly \emph{increasing} sound speeds.

There are a number of outstanding questions that lend themselves to further work. Most noteworthy is that, in this paper, we considered the local-type contribution to $\fnl$ from multiple-field dynamics during inflation. It would therefore be interesting to compare this to the equilateral contribution produced at horizon exit by  small sound speeds. Whilst this contribution is dominant in the single field case, the introduction of multiple fields may alter this conclusion via conversion between entropy and adiabatic modes (see, for example, \cite{Langlois08,Langlois08b,Arroja08}). Moreover, correlations between different contributions may lead to one of the few explicit examples of models characterised
by combined equilateral and local non-Gaussianities (see \cite{RenauxPetel09} for an alternative). We intend to address these issues in much greater detail in a follow-up paper. 

Other open questions remain regarding the application to more general cases. Firstly, whilst our example trajectory illustrates the production of local-type non-Gaussianity through non-trivial sound speeds, it is unclear how generic this behaviour is. It would therefore prove useful to study a wider parameter space (see \cite{Byrnes08} for an example in the canonical case). Similarly, our use of a sum separable Hubble parameter, whilst providing analytically tractable expressions, also restricts the choice of potential. Thus it would be interesting to study more general cases with alternative analytical or numerical methods. Note that this is also true in the canonical case. Finally, we highlight that our model based approach explicitly demonstrates the dynamics necessary for local-type non-Gaussianity. This must be considered in light of more general methodologies however, such as an effective field theory approach \cite{Cheung08,Senatore10}, which offer wider applicability. It would therefore be interesting to bridge the gap between the two. 

We end by again noting that our approach treats multiple-field dynamics \emph{during} inflation and, since we did not consider a model of reheating or an approach to adiabaticity, we cannot conclude that such values are necessarily observed in the CMB. In our illustrative example however, the non-linearity parameter peaks at the end of inflation, corresponding to a violation of slow variation at the tips of the throats. We would argue that such scenarios seem the most likely to produce non-Gaussianity during inflation that can imprint upon the CMB. More generally however, it would be interesting to apply similar techniques to \cite{Elliston11a,Elliston11b,Meyers11-a,Meyers11-b,Peterson11a,Watanabe11,Choi12,Mazumdar12} in the case of non-standard kinetic terms, to further understand what we can infer about the dynamics of inflation from observations of non-Gaussianity.

\acknowledgments

The authors would like to thank Taichi Kidani, Kazuya Koyama and Mathieu Pellen for useful discussions and Guido W. Pettinari for technical help. We also thank the anonymous referee for their helpful comments. JE is supported by an STFC PhD studentship. GT is supported by an STFC Advanced Fellowship ST/H005498/1. DW is supported by STFC grant ST/H002774/1.

\appendix

\section{Equations of motion in the Hamilton-Jacobi formalism\label{sec:appendix}}

\noindent In this appendix we derive the equations of motion (\ref{eq:field-general}) and (\ref{eq:friedman-general}), presented in section~\ref{sec:model-homogeneous}, in Hamilton-Jacobi form. This essentially extends the calculation in \cite{Salopek90} to non-canonical cases satisfying the following action
    
      \vspace{5pt}
      \be S=\frac{1}{2} \int d^{4}x\sqrt{-g}\bigl[R+2\sum_{I}P_{I}-2V\bigr],\label{eq:action-general-app}\ee
      \vspace{5pt}
    
\noindent  where $R$ is the Ricci scalar and $g$ is the determinant of the metric tensor $g_{\mu\nu}$. Note that $P_{I}$ is a function of the single scalar field $\phi_{I}$ and kinetic function $X_{I}=-\frac{1}{2}g^{\mu\nu}\phi_{I,\mu}\phi_{I,\nu}$, whereas the potential $V$ is a function of the set of scalar fields $\phi=\{ \phi_{1},\phi_{2},...,\phi_{N}\}$. Note also that we return to the canonical case with $P_{I}=X_{I}$. 

It is useful to recast the action (\ref{eq:action-general-app}) using the Arnowitt-Deser-Misner (ADM) formalism \cite{Arnowitt62}, originally conjectured as a Hamiltonian formulation of general relativity. The formalism proves useful for studying the background evolution as it naturally produces scalar field equations in Hamilton-Jacobi form \cite{Salopek90,Kinney97}, which is more suited to non-trivial sound speeds. Moreover, the formulation introduces the lapse function $N$ and shift vector $N^{i}$ which under variation become Lagrange multipliers, simplifying the equations of motion for perturbations. The ADM metric is given by
  
      \vspace{3pt}
      \be ds^{2}=-N^{2}dt^{2}+h_{ij}\left(dx^{i}+N^{i}dt\right)\left(dx^{j}+N^{j}dt\right),\ee
      \vspace{3pt}

\noindent where $h_{ij}$ is the spatial 3-metric. In terms of this metric, the action (\ref{eq:action-general-app}) and kinetic term becomes

      \vspace{-10pt}
      \begin{align}	
      S&=\frac{1}{2} \int d^{4}x\sqrt{h}\bigl[NR^{(3)}+NK_{ij}K^{ij}-NK^{2}+2N\sum_{I}P_{I}-2NV\bigr],\label{eq:action_app}\\[7.5pt]
      X_{I}&=\frac{1}{2N^{2}}\left(\dot{\phi}_{I} - N^{i}\phi_{I,{i}}\right)^{2}-\frac{1}{2}\phi_{I,{i}}\phi_{I}^{,{i}},
      \end{align}
      \vspace{0pt}

\noindent where $K=K^{i}_{i}$, $R^{(3)}$ is the three dimensional Ricci scalar and indices are raised and lowered using the spatial metric. $K_{ij}$ is the extrinsic curvature, given by

      \vspace{0pt}
      \be K_{ij}=\frac{1}{2N}\left(N_{i|j}+N_{j|i}-\dot{h_{ij}}\right),\ee
      \vspace{0pt}

\noindent  where $_{|i}$ denotes the covariant derivative with respect to the spatial metric. It is convenient to define the traceless part of a tensor with an overbar and we use the extrinsic curvature as an example

      \vspace{-6pt}
      \be \bar{K}_{ij}=K_{ij}-\frac{1}{3}h_{ij}K.\ee
      \vspace{-1pt}

\noindent It is trivial to check that $\bar{K}_{ij}$ is indeed traceless. Variation of the action (\ref{eq:action_app}) with respect to $N$ and $N^{i}$ yields the energy and momentum constraints

      \vspace{-10pt}
      \begin{align}
      \bar{K}_{ij}\bar{K}^{ij}-\frac{2}{3}K^{2}-R^{(3)}+2\rho=0,\label{eq:energy-constraint-app} \\[10pt]
      \bar{K}^{j}_{i|j}-\frac{2}{3}K_{|i}+\sum_{I}\Pi_{I}\phi_{I,{i}}=0.\label{eq:momentum-constraint-app}
      \end{align}

\noindent Notice that these relations are identical in form to the canonical results of \cite{Salopek90}, since the effect of the non-canonical kinetic terms are packaged into the energy density $\rho$ and momenta $\Pi_{I}$ of the fields

      \vspace{-15pt}
      \begin{align} 
      \rho&=\sum_{I}\left(\frac{\Pi_{I}^{2}}{P_{I,X_{I}}}-P_{I}\right)+V,\\[10pt]
      \Pi_{I}&=\frac{P_{I,X_{I}}}{N}\left(\dot{\phi}_{I}-N^{i}\phi_{I,{i}}\right).
      \end{align}
      \vspace{0pt}

\noindent  Similarly, variation of (\ref{eq:action_app}) with respect to the spatial metric yields the gravitational field equations

      \vspace{-17.5pt}
      \begin{align}
      &\dot{K}-N^{i}K_{|i}+N^{|i}_{|i}-N\left(\frac{3}{4}\bar{K}_{ij}\bar{K}^{ij}+\frac{1}{2}K^{2}+\frac{1}{4}R^{(3)}+\frac{1}{2}S\right)=0,\label{eq:trace-app}\\[10pt]
      & \dot{\bar{K}}^{i}_{j}+N^{i}_{|k}\bar{K}^{k}_{j}-N^{k}_{|j}\bar{K}^{i}_{k}-N^{k}\bar{K}^{i}_{j|k}+N^{|i}_{|j}- \nonumber \\[5pt]  & \hspace{112pt} \frac{1}{3}N^{|k}_{|k}\delta^{i}_{j}-N\left(K\bar{K}^{i}_{j}+\bar{R}^{(3)i}_{\phantom{(3)}j}-\bar{S}^{i}_{j}\right)=0,
      \end{align}

\noindent where the stress three tensor is defined as

      \be S_{ij}=\sum_{I}\Bigl(P_{I,X_{I}}\phi_{I,{i}}\phi_{I,{j}}+h_{ij}P_{I}\Bigr)-h_{ij}V.\label{eq:stress-app}\ee

\noindent Again the non-canonical contributions are packaged into (\ref{eq:stress-app}). Finally, the field equations follow from variation with respect to $\phi_{I}$

      \vspace{-10pt}
      \be NP_{I,I}-NV_{,I}-\frac{1}{\sqrt{h}}\dot{\left(\sqrt{h}P_{I,X_{I}}\Pi_{I}\right)}+\Bigl(P_{I,X_{I}}N^{i}\Pi_{I}\Bigr)_{|i}+\left(NP_{I,X_{I}}\phi_{I}^{,{i}}\right)_{|i}=0.\ee
      \vspace{1pt}
	  
\noindent In the canonical limit the above system of equations recover the results of \cite{Salopek90}. 

To derive a set of equations for the homogeneous evolution in Hamilton-Jacobi form we again follow \cite{Salopek90} by expanding in spatial gradients. We consider only terms up to first order in spatial gradients (e.g. neglecting terms such as $N^{|i}_{|i}$ and $\bar{R}^{(3)i}_{\phantom{(3)}j}$) and set $N^{i}=0$  in order to simplify the system of equations. The analysis follows \cite{Salopek90} very closely and so for brevity we simply quote the results. The momentum constraint (\ref{eq:momentum-constraint-app}) becomes

      \vspace{2pt}
      \be H_{,i}=-\frac{1}{2}\sum_{I}\Pi_{I}\phi_{I,{i}},\label{eq:intermediate-result-app}\ee
      \vspace{2pt}

\noindent where we have replaced the trace of the extrinsic curvature with the spatially dependent Hubble parameter $H(t,x^{i})=-\frac{K(t,x^{i})}{3}$. Since $H_{,i}=\sum_{I}H_{,I}{\phi_{I,{i}}}$, matching terms with (\ref{eq:intermediate-result-app}) and considering a spatially flat Friedmann-Robertson-Walker Universe (such that $N=1$) gives the field equations

			\vspace{-5pt}
      \be \dot{\phi}_{I}=-\frac{2}{P_{I,X_{I}}}H_{,I}.\ee 
      \vspace{7pt}

\noindent The Friedmann equation follows directly from either the energy constraint (\ref{eq:energy-constraint-app}) or the equation for the trace of the extrinsic curvature (\ref{eq:trace-app})

      \vspace{4pt}
      \be 3H^{2}=\sum_{I}\left(\frac{4H_{,I}^{2}}{P_{I,X_{I}}}-P_{I}\right)+V.\ee
      \vspace{4pt}

\noindent Finally then, we arrive at the equations of motion (\ref{eq:field-general}) and (\ref{eq:friedman-general}) used in the beginning of section~\ref{sec:model-homogeneous}.

\end{document}